\documentclass[compsoc, conference, letterpaper, 10pt, times]{IEEEtran}
\usepackage{booktabs}
\usepackage[T1]{fontenc} 
\usepackage{amsmath}
\usepackage[colorinlistoftodos,prependcaption,textsize=tiny]{todonotes}
\usepackage{hyperref}

\usepackage{mathtools}
\usepackage{url}
\usepackage[noend]{algpseudocode}
\usepackage{tabularx,colortbl}
\usepackage{multirow}
\usepackage[Symbolsmallscale]{upgreek}
\usepackage{algorithm}
\usepackage{mathtools}
\usepackage{url}
\usepackage[noend]{algpseudocode}
\usepackage{siunitx}
\usepackage{listings}
\usepackage{gensymb}
\usepackage{color,soul}

\begin{document}

\date{}

\title{Piping Botnet - Turning Green Technology into a Water Disaster}


\author{

Ben Nassi, Moshe Sror, Ido Lavi, Yair Meidan, Asaf Shabtai, Yuval Elovici\\
\{nassib,srorm,laid,yairme,shabtaia,elovici\}@post.bgu.ac.il\\
\textbf{Video}  \url{https://www.youtube.com/watch?v=Yy8tOEhH6T0}\\


\and
} 
\maketitle

\thispagestyle{empty}

\section*{Abstract}

The current generation of IoT devices is being used by clients and consumers to regulate resources (such as water and electricity) obtained from critical infrastructure (such as urban water services and smart grids), creating a new attack vector against critical infrastructure. In this research we show that smart irrigation systems, a new type of green technology and IoT device aimed at saving water and money, can be used by attackers as a means of attacking urban water services. We present a distributed attack model that can be used by an attacker to attack urban water services using a botnet of commercial smart irrigation systems. Then, we show how a bot running on a compromised device in a LAN can: (1) detect a connected commercial smart irrigation system (RainMachine, BlueSpray, and GreenIQ) within 15 minutes by analyzing LAN's behavior using a dedicated classification model, and (2) launch watering via a commercial smart irrigation system according to an attacker's wishes using spoofing and replay attacks. In addition, we model the damage that can be caused by performing such an attack and show that a standard water tower can be emptied in an hour using a botnet of 1,355 sprinklers and a flood water reservoir can be emptied overnight using a botnet of 23,866 sprinklers. Finally, we discuss countermeasure methods and hypothesize whether the next generation of plumbers will use Kali Linux instead of a monkey wrench.

\section{Introduction}
A variety of IoT devices are being deployed across cities around the world as part of the smart city trend. Hundreds of cities in Europe, Asia, Australia, America, and even Africa, have already adopted smart city technologies and use them to obtain information that helps them manage assets and resources efficiently \cite{Top-10-smart-cities-1,smart-africa}. IoT devices are currently used by consumers and clients to regulate and monitor resources obtained from critical infrastructure including energy, water, etc. The interface between an IoT device with Internet connectivity (which is located on the consumer side) and the cyber-physical system (CPS) of critical infrastructure (which is located on the provider side) necessitates that all of the connected links between the two meet the most rigorous security standards. Cyber attacks targeted at critical infrastructure may result in urban disaster as happened in the cyber attack against Ukraine's power grid which left 700,000 people without electricity for several hours \cite{cyberattack-on-Ukrainian-power-grid}. In order to prevent attackers from attacking the networks and physical systems of critical infrastructure, various steps are taken to secure these systems, including: (1) buying equipment, hardware, and systems from trusted parties, (2) deploying an= security solution such as IDS/IPS, and (3) physically disconnecting the networks from the Internet (air-gapping their networks). While critical infrastructures minimize any possible attack vector against them, IoT devices with Internet connectivity that are located on the consumer side (e.g., in smart homes, smart cities, etc.) and used to regulate a resource obtained from the critical infrastructure remain the weakest link in this interface.
Such IoT devices have created a new attack vector for critical infrastructure and will soon become a prime target for attackers.

In this paper, we show how critical infrastructure that adheres to very strict security standards can be attacked indirectly using a botnet of IoT devices with Internet connectivity. We demonstrate how an attacker can exploit IoT devices, which are deployed across a smart city and used to regulate a resource obtained via the CPS of critical infrastructure, as a means of attacking the critical infrastructure. To make our discussion about this type of attack more concrete, we focus on a new type of IoT device: smart irrigation systems. Considered a green technology, smart irrigation systems are a good target for analysis because: (1) they have already been adopted by smart cities (e.g., Barcelona \cite{barcelona-irrigation-1}), agriculture, and the private sector around the world, (2) they regulate water flow for watering and irrigation, a resource which is provided by urban water services  (critical infrastructure), and (3) they are considered a key actor in the smart water grid revolution, because they are aimed at automating water irrigation in order to save water and will soon replace most traditional irrigation systems.

First, we present a distributed attack model that can be used by an attacker to attack urban water services using a botnet of commercial smart irrigation systems (Section \ref{section:Adversarial Attack Model}). Then we show how a bot running on a compromised device in a LAN can: (1) detect a connected commercial smart irrigation system (RainMachine \cite{RainMachine}, BlueSpray \cite{BlueSpray}, and GreenIQ \cite{GreenIQ}) within 15 minutes by analyzing LAN's traffic behavior using a dedicated classification model (Section \ref{section:Detecting smart irrigation systems}), and (2) launch watering via a commercial smart irrigation system according to an attacker's wishes using spoofing and replay attacks (Sections \ref{section:MiTM} and \ref{section:replay-attacks}). In addition, we model the damage that can be caused by performing such an attack (Section \ref{section:damage}) and show that a standard water tower can be emptied in an hour using a botnet of 1,355 sprinklers and a flood water reservoir can be emptied overnight using a botnet of 23,866 sprinklers. We also discuss countermeasure methods.

In this research, we make the following contributions: (1) while previous attacks against critical infrastructure required the attacker to compromise the systems of critical infrastructure, we present an attack against critical infrastructure that does not necessitate compromising the infrastructure itself and is done indirectly by attacking attacking client infrastructure that is not under the control of the critical infrastructure provider. In addition, we show that a bot running on a compromised device can (2) detect a smart irrigation system connected to its LAN in less than 15 minutes, and (3) launch watering via the smart irrigation system using various methods.

\section{Related Work}
\label{section:Related Work}

In this section we describe related work on attacks against critical infrastructure and provide an overview of DDoS attacks using IoT devices. \textbf{Critical infrastructure} has been defined by the European Commission as an "\textit{asset or system which is essential for the maintenance of vital societal functions}" \cite{Critical-infrastructure-EU}. The Department of Homeland Security identifies 16 sectors as critical infrastructure including water/wastewater systems, energy, nuclear reactors, chemical, dams, emergency services, etc. \cite{DHS-Critical-Infrastructure-Sectors}. Some of these sectors provide 24/7 services, while others regulate continuous real-time processes using dedicated cyber-physical systems such as controllers, sensors, etc. Definitions mentioned in the literature for a CPS include a networked/distributed control system (NCS/DCS), sensor actuator network (SAN), wireless industrial sensor network (WISN), industrial control system (ICS), and supervisory control and data acquisition (SCADA) networks \cite{5560697}. In the remainder of this article we will refer to such systems as CPSs.

The interest of adversaries in attacking a CPS of critical infrastructure began three and a half decades ago. The first known cyber attack was launched in 1982 by intruders who planted a Trojan in the SCADA system that controls the Siberian pipeline and caused an explosion equivalent to three kilotons of TNT \cite{Miller:2012:SSC:2380790.2380805}. In recent years there has been a significant increase in the number of cyber attacks against critical infrastructure \cite{ponemon-report} which can even result in death \cite{Miller:2012:SSC:2380790.2380805}. Two famous cyber attacks against critical infrastructure that were launched during the last 10 years and resulted in a large amount of damage are the cyber attack against Ukraine's power grid which left 700,000 people without electricity for several hours \cite{cyberattack-on-Ukrainian-power-grid}, and Stuxnet which was targeted at Iran's nuclear plant and caused a large number of centrifuges to be taken offline \cite{stuxnet}. 

Air-gapping (isolating the networks from the Internet) is typically applied to systems of critical infrastructures in order to prevent attackers from compromising CPSs via the Internet. Air-gapping requires the attackers to physically compromise the critical infrastructure in order to attack it. However, motivated attackers use various attack vectors to compromise critical infrastructures using: (1) supply chain attacks \cite{urciuoli2013supply}, (2) innocent or malicious insiders, and (3) social engineering. Many methods \cite{mitchell2014survey, cheung2007using} to detect and mitigate cyber attacks against the CPSs of critical infrastructure have been suggested over the years, and security tools such as IDSs/IPSs are used for this purpose. Recently, several studies raised concerns regarding the cyber security of existing  and future critical infrastructure (e.g., the smart grid) \cite{6016202, 6141833, 6257525, wang2013cyber}, while other studies have specifically discussed using IoT devices as an attack vector to disrupt the operation of critical infrastructure \cite{simon2017critical, 7959707, BEKARA2014532, sajid2016cloud}.

\textbf{DDoS attacks} are DoS attacks that are usually launched from a group of compromised devices (botnet); each device (bot) in the botnet is infected with a malicious agent. A wide range of studies have been published on this subject, demonstrating different types of DDoS attacks \cite{kuhrer2014hell} and related detection \cite{sekar2006lads} and defense \cite{fayaz2015bohatei} techniques. DDoS attacks are considered one of the major threats and most challenging problems of today's cyber security world \cite{zargar2013survey}. In this overview we focus specifically on distributed attacks that rely on IoT devices as bots or targets. The earliest known IoT botnet is Linux/Hydra \cite{janus2011heads, de2017analysis}, which was released in 2008, and specifically aimed at routing devices based on MIPS architecture \cite{de2017analysis}. Since 2008, many types of IoT botnets have appeared in the wild \cite{de2017analysis}. Probably the most famous IoT botnet is Mirai, which turned a large number of IP cameras running the Linux OS into remote controlled bots that were used to launch a massive DDoS attack in 2016 \cite{203628,7971869}. Variants of the Mirai botnet were found to be used in attacks against different targets during 2017 \cite{7971869}. 

\begin{figure*}
\centering
\includegraphics[width=0.62\textwidth]{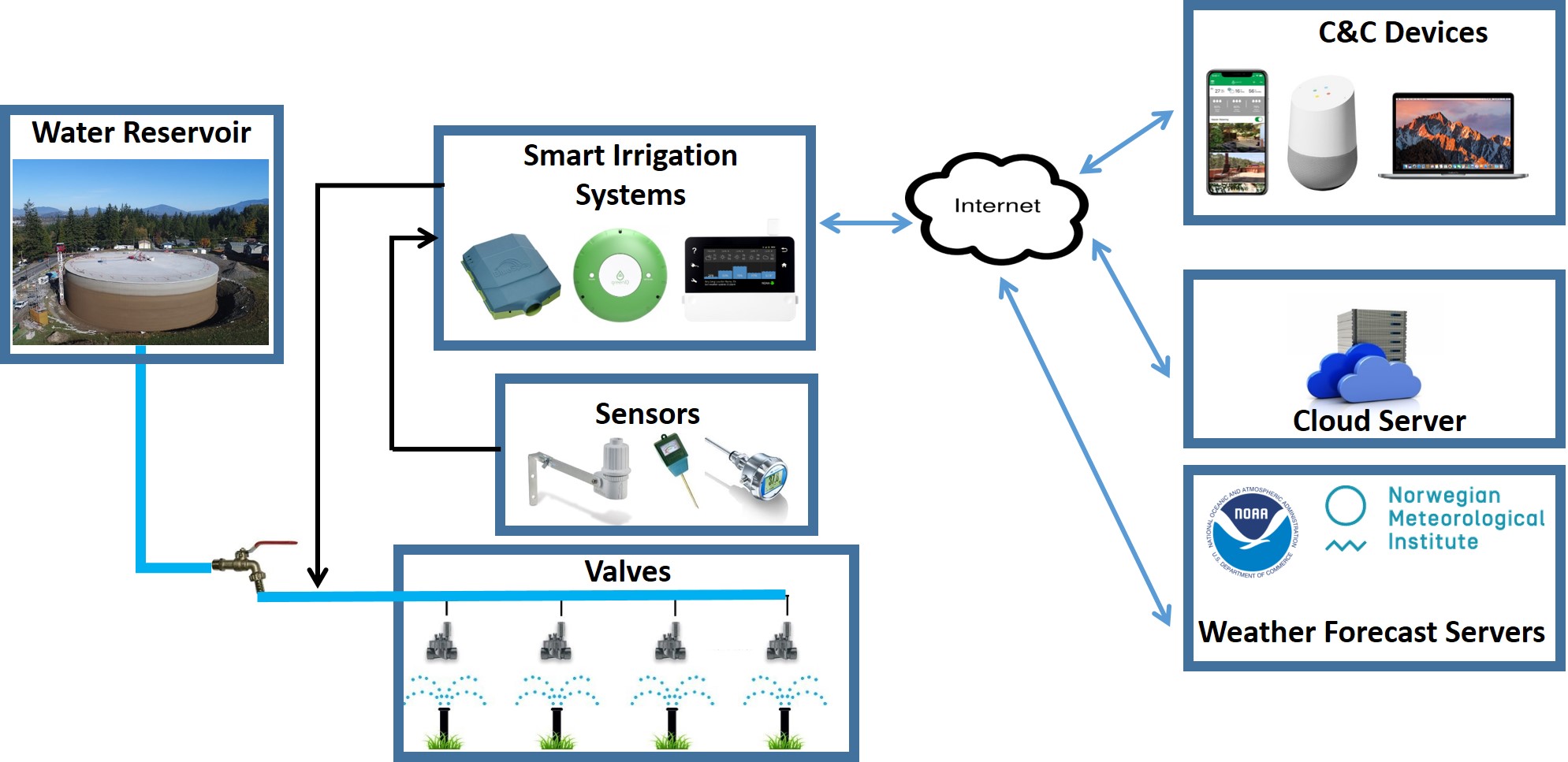}
\caption{Smart irrigation systems regulate watering by consuming water from the urban water service and interface with various sensors, weather forecast services, C\&C devices, and dedicated cloud servers.}
\label{fig:eco-system}
\end{figure*}

Usually, IoT botnets exploit vulnerabilities in operating systems and protocols in order to compromise devices and launch their attack \cite{de2017analysis}. Recently new kinds of DDoS attacks have been introduced. A recent study \cite{guri20179} described a TDoS (telephony denial of service) attack against 911 emergency services in which many calls to the service were triggered simultaneously. The TV was used twice as a means of launching a distributed attack on smart assistants' servers, intentionally via a Burger King advertisement \cite{burger-king-against-google-home} and accidentally via a daily news program \cite{alexa-order-dollhouse}. In both cases, a voice command that was produced from the TV triggered many smart assistants to launch a large number of requests to their servers at the same time. 

To the best of our knowledge, we are the first to (1) present a distributed attack against critical infrastructure that does not require compromising its systems, and (2) create a botnet that uses smart irrigation systems connected to the Internet as means of attacking critical infrastructure. 

\section{Smart Irrigation Systems}
\label{section:Smart irrigation system}

Smart irrigation systems refer to \textit{advanced irrigation systems that incorporate various sensors and network components for better efficiency} \cite{smart-irrigation-systems-market}. Smart irrigation systems, a new type of green technology and IoT device, are equipped with Internet connectivity that facilitates communication with sensors, weather forecast services, C\&C devices, and dedicated cloud servers. The prime motive behind the advent of smart irrigation systems is to enhance the overall water efficiency of irrigation systems, with minimal user effort. Internet connectivity is designed to provide remote access capabilities via any device (mobile, personal computer, etc.) and automatic adjustment of water consumption based on data that is retrieved from weather forecast services, without any manual interaction.

Smart irrigation systems use Internet connectivity for various operations (e.g., automatic watering regulation, remote C\&C, etc.). Most smart irrigation systems provide Wi-Fi communication via an integrated NIC, however there are some smart irrigation systems that provide GSM communication via a GSM dongle with a SIM card (as can be seen in Shodan's results when performing a search for the word "BlueSpray"\footnote{\label{Shodan-BlueSpray} \url{https://www.shodan.io/search?query=bluespray}}). Smart irrigation systems are physically connected to a set of valves that are connected to the main water line on one end and to pipelines/sprinklers on the other end. The valves are controlled by the smart irrigation system and used to adjust the water flow from the main water line to sprinklers and droppers. The valves that smart irrigation systems can control independently are called zones, and the number of zones varies from 4 to 24 in most typical systems.

Smart irrigation systems were first introduced in 2013, and in the next few years they will replace most traditional irrigation systems around the world because (1) they are inexpensive (their price starts at \$100) and designed to save money and water, (2) they provide a convenient remote HMI for C\&C via smartphones, smart assistants, and computers, in contrast to traditional irrigation systems which have a dedicated display, and (3) they monitor water consumption and present the watering history. In addition to all of the abovementioned reasons, smart irrigation systems will replace traditional irrigation systems, because they were identified by \cite{mutchek2014moving, YE2016361, doi:10.1080/19443994.2014.917887} as a key actor in the future smart water grid architecture, which is referred to as a \textit{real-time two-way network equipped with sensors and devices for continuous and remote monitoring of the water distribution system} \cite{Smart-Water-Grid-Market}. 

Smart irrigation systems are designed to support the following functionality: (1) provide remote HMI communication (for purposes of scheduling a watering plan, presenting the watering history, etc.) over the Internet to C\&C devices (using a dedicated application for smartphones, a Web user interface for browsers, and a voice interface for smart assistants), (2) monitor water consumption, and (3) automatically adapt the watering plan according to data that is obtained from weather forecast services (e.g., precipitation forecast for the next few days) and sensors (e.g., obtain information regarding soil moisture).


\begin{figure*}[tbp]
\centering
\includegraphics[width=0.75\textwidth]{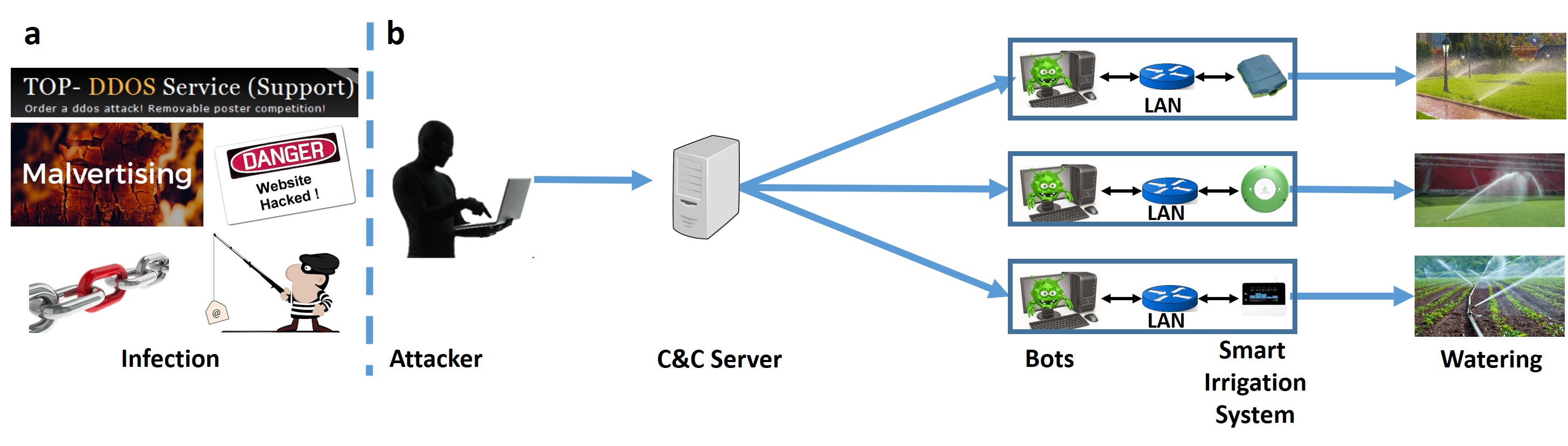}
\caption{The adversarial attack model.}
\label{fig:attack_model}
\end{figure*}

Figure \ref{fig:eco-system} outlines the entire smart irrigation system ecosystem. As can be seen in Figure \ref{fig:eco-system}, smart irrigation systems typically interface with the following  entities:

1) \textbf{Weather Forecast Service} - There are many weather forecast services on the Internet \cite{NOAA, Metno, wunderground, FAWN, CIMIS, DarkSky,PwsWeather} that provide a REST API in which a request that contains the location of the desired weather forecast is sent from a client and followed by a response from the weather forecast service that contains the weather forecast (temperature, humidity, wind direction, wind speed, pressure, cloudiness, etc.) for each hour/part of day for the upcoming days/week. Smart irrigation systems use weather forecasts in order to adjust their watering plan and typically launch a few requests a day to obtain updates.

2) \textbf{C\&C Device} - Smart irrigation systems provide an HMI for C\&C that is based on a Web browser, mobile/tablet application, and smart assistants. The HMI provides smart irrigation system users with various capabilities to remotely control and monitor the operation of smart irrigation systems from anywhere around the world (e.g., to schedule a watering program, to visualize weekly aggregated watering consumption data) using a cloud server that mediates between the C\&C device and the smart irrigation system. This is a very convenient interface compared to that of traditional irrigation systems which don't provide a remote or visual HMI for C\&C.

3) \textbf{Cloud Server} - Each smart irrigation system communicates with its own cloud server. The primary role of the cloud server is to mediate between the C\&C device and the smart irrigation system. In addition, the cloud server also provides a firmware update, stores the smart irrigation system's configuration, and stores the watering history. Smart irrigation systems typically launch an update request that contains their identifier to the cloud server once a minute in order to verify whether new updates have been sent from the user.

4) \textbf{Sensors} - Smart irrigation systems provide a
wired/wireless interface for sensors (e.g., precipitation, soil moisture, temperature, and water flow sensors). Based on the data that is obtained from the connected sensors, smart irrigation systems adjust the watering plan and regulate their operation.

In this research we analyzed three commercial smart irrigation systems: RainMachine \cite{RainMachine}, BlueSpray \cite{BlueSpray}, and GreenIQ \cite{GreenIQ} that were identified as three of the 10 most advanced smart irrigation systems by \cite{top-sprinklers-1} and \cite{top-sprinklers-2}. They contain up to 24 valves, and they are able to communicate with sensors (e.g., precipitation sensor) and weather forecast services.



\section{Adversarial Attack Model}
\label{section:Adversarial Attack Model}

In this section we describe the attacker's threat model.
We consider an \textbf{attacker}, a malicious entity, that applies a distributed attack on the urban water service using a botnet of smart irrigation systems in order to cause harm to society. The attacker's objective can be any one of the following:

1) \textbf{To Waste Water} - usually, water is purified in a treatment plant after it has been pumped from a natural water source (e.g., groundwater). From the treatment plant, the water is distributed to urban/areal reservoirs and tanks that distribute water for residents in the entire distribution area. In some places, areal reservoirs and water tanks are not physically connected to a treatment plant using pipelines due to physical limitations. Instead, areal reservoirs are filled with water shipped to the reservoir on a weekly/monthly basis or when the reservoir is nearly empty. Applying an attack that wastes water and empties the urban water reservoir may result in the inability to provide water to residents until the local water reservoir can be refilled. In addition, in many places around the world, there is a serious water shortage \cite{36-Most-Water-Stressed-Countries}, so wasting water is even more dangerous.

2) \textbf{Financial Damage} - attacking smart irrigation systems increases water consumption and causes financial loss to cities that use irrigation systems to water parks and private households that use irrigation systems for watering their yard/garden. In many places around the world, water is expensive. For example, the average combined water tariff in Portland, Oregon is \$8.00 per cubic meter of water \cite{water-tariff}.

3) \textbf{Reducing Water Flow} - by applying a distributed attack against many smart irrigation systems that are connected by the same pipeline to the urban water service, the attacker can also reduce water flow in all of the households connected to the pipeline.

We do not consider a targeted attack against a specific smart irrigation system (e.g., attacking a neighbor) a dangerous attack, because the result of such an attack is limited to financial damage to one user (as a result of wasting water). In contrast, we consider an attack that is directed at an urban/local water service a very dangerous attack, because preventing people from accessing a resource from critical infrastructure can be a disaster \cite{cyberattack-on-Ukrainian-power-grid}, depending on the number of clients affected and prevented from accessing the resource. Attacking urban water services requires the attacker to use many smart irrigation systems. Figure \ref{fig:attack_model} presents the attacker's threat model. A botnet is used by the attacker to launch massive water consumption by many smart irrigation systems simultaneously. The attack consists of three stages:

\textbf{Stage 1 - infection}: the attacker builds a botnet of smart irrigation systems. The attacker can rent botnet services \cite{botnets-for-rent-1,botnets-for-rent-2} which are traded for bitcoin on the darknet. Alternatively, the attacker can infect devices that are connected to the Internet (e.g., laptop, smartphone, router, etc.) with malware using common infection vectors (e.g., email attachments, compromised websites, malvertising campaigns, and supply chain attacks), as can be seen in Figure \ref{fig:attack_model}a.

\textbf{Stage 2 - reconnaissance}: each bot searches for smart irrigation systems that are connected to its LAN. If no connected smart irrigation systems are found, the bot destroys itself in order to cover its tracks. In Section \ref{section:Detecting smart irrigation systems} we show that a smart irrigation system can be detected in a LAN within 15 minutes by a bot running on a device connected to the same LAN by analyzing outgoing traffic.

\textbf{Stage 3 - attack}: at the appropriate time, the attacker signals the botnet to apply a distributed attack that results in massive water consumption of the urban water service. The attacker uses the bots to attack the smart irrigation systems connected to their LANs using various attack vectors (we describe them in Sections \ref{section:MiTM} and \ref{section:replay-attacks}) that cause high water consumption, as can be seen in Figure \ref{fig:attack_model}b.

In order to coordinate the DDoS attack, the attacker communicates with the bots using a C\&C server. A common C\&C approach is based on managing bots via one or more C\&C servers located somewhere on the Internet. The IP addresses or domain names of the C\&C server are hidden in the bots' code and may be updated later via the C\&C. Upon installation, a bot connects to its C\&C server over a secure network protocol (e.g., HTTPS) and receives commands. The botnet operator notifies the C\&C servers to send either a START or STOP command to the bots. A START command will contain parameters such as the start time and the duration. Optionally, a location can be provided to ensure that only bots located in a certain geographical region are activated in order to focus on a specific urban water service. Additional C\&C mechanisms for botnets can be found in \cite{zeidanloo2009botnet}. It is important to note that there are more advanced topologies which are resilient to being shut down, e.g., peer-to-peer, hierarchical, and random topologies.

In Sections \ref{section:Detecting smart irrigation systems} and \ref{section:MiTM} we show how a bot can (1) detect smart irrigation systems in its LAN within 15 minutes, and (2) control a smart irrigation system using various attacks.

\section{Analysis and Reverse Engineering}
\label{section:analysis}

\begin{figure}
\centering
\includegraphics[width=0.75\columnwidth]{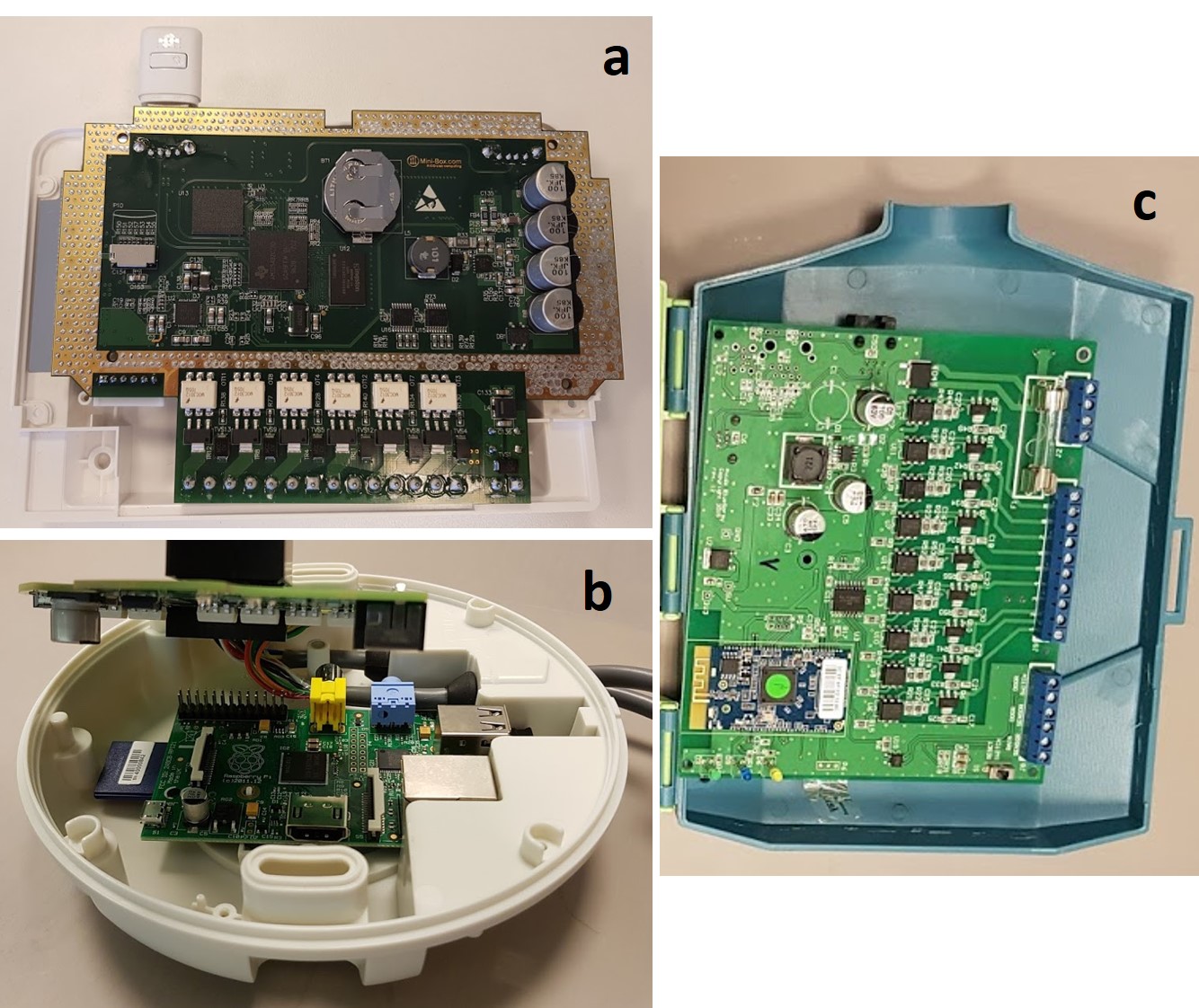}
\caption{SoC board of commercial irrigation systems: (a) RainMachine \cite{RainMachine}, (b) GreenIQ \cite{GreenIQ}, and (c) BlueSpray \cite{BlueSpray}. The GreenIQ smart irrigation system contains a Raspberry Pi as the SoC.}
\label{fig:microcontrollers}
\end{figure}

In this section we describe the analysis that we performed for three commercial smart irrigation systems (GreenIQ, RainMachine, and BlueSpray). We combined two techniques: (1) we connected all three smart irrigation systems to a router and captured their ingoing/outgoing traffic for a few days. We analyzed their connections with their C\&C devices, cloud servers, and weather forecast services from the captured PCAP files using Wireshark. In addition, (2) we reverse engineered commercial smart irrigation systems by extracting their firmware. The GreenIQ second generation smart irrigation system is based on a Raspberry Pi controller board with a connected SD card (as can be seen in Figure \ref{fig:microcontrollers}b). We copied the content of the SD card to a laptop using an SD card reader and found 34 Python files that the firmware is based on. Unlike the GreenIQ smart irrigation system which uses a Raspberry Pi as its controller board, RainMachine does not use a commercial board and designed its own controlling circuitry. We used a USB to UART adapter (FT232R) to extract RainMachine's firmware from the SoC's UART terminals, a technique that was shown in \cite{DBLP:conf/cardis/ShwartzMBEO17}. RainMachine runs a modified version of the Android OS, so we looked for the APK of RainMachine's application and found the file \textit{RainMachine-UI.apk}. We extracted the APK to Java files using an online decompiler tool. The firmware of GreenIQ and RainMachine was not obfuscated.





\section{Detecting Connected Smart Irrigation Systems}
\label{section:Detecting smart irrigation systems}

During the reconnaissance stage, each bot must detect whether a smart irrigation system is connected to its LAN. If no smart irrigation system is found, then the bot sends a notification to the C\&C server and destroys itself in order to cover its tracks. We decided to design and empirically evaluate a model that detects a connected smart irrigation system and is used by a bot running on a compromised device which is connected to the same LAN (of the smart irrigation system); detection is based on analyzing the captured network traffic data of suspicious IP. In order to do so, we connected three commercial smart irrigation systems (RainMachine~\cite{RainMachine}, BlueSpray~\cite{BlueSpray}, and GreenIQ~\cite{GreenIQ}) to a router via Wi-Fi, and monitored the LAN traffic using a bot that was installed on a laptop that was connected to the same LAN (by applying ARP spoofing from the laptop to the smart irrigation systems). We extracted several features from the captured traffic data, and appended another set of traffic data with the same features (collected from various IoT devices in other research \cite{Meidan:2017:PML:3019612.3019878}) to them. The IoT data was obtained from numerous and various IoT device types that can be found in standard homes nowadays: two smart bulbs, a smart refrigerator, sixteen security camera, two laptops, two smartphones, and five smartwatches. 

\begin{figure}
\centering
\includegraphics[width=0.8\columnwidth]{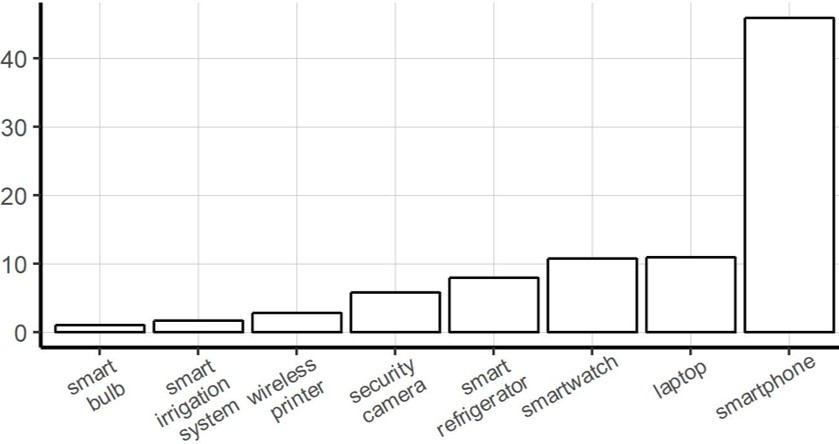}
\caption{The  average number of unique destinations that IoT devices interface with in an hour.}
\label{fig:destinations}
\end{figure}

In our preliminary analysis, we explored the average number of unique destinations that smart irrigation systems interface with per hour, and compared the results with the abovementioned IoT devices. As can be seen in Figure \ref {fig:destinations}, the average number of unique destinations that smart irrigation systems interface with is very low compared with the smartphones and smart refrigerator. However, a small average number of unique destinations is a property that is common to most of the IoT devices we analyzed so it cannot be used by a bot to determine whether a suspicious IP is a smart irrigation system or not. 

Following this preliminary analysis, we looked for unique characteristics that could be used by a bot running on a LAN to decide whether a connected device is a smart irrigation system or not. Currently, the manufacturers of smart irrigation systems do not produce any other types of IoT devices \cite{top-sprinklers-1,top-sprinklers-2}. With this observation in mind, we decided to analyze the identity of the cloud servers that smart irrigation systems interface with. Unlike Samsung's cloud server which supports many IoT devices manufactured by Samsung (smart refrigerator, smartphone, etc.), the cloud servers of the tested smart irrigation systems interface only with their respective smart irrigation systems. A packet sent to GreenIQ cloud server cloud server was sent only from GreenIQ smart irrigation system. The same thing is also true for BlueSpray and RainMachine during the 26 hour period of data collection. Hence, due to the absence of overlap between the contacted cloud servers, an outgoing packet sent to a smart irrigation system cloud server can clearly and reliably indicate that the packet's sender is a smart irrigation system.

\begin{figure}
\centering
\includegraphics[width=0.8\columnwidth]{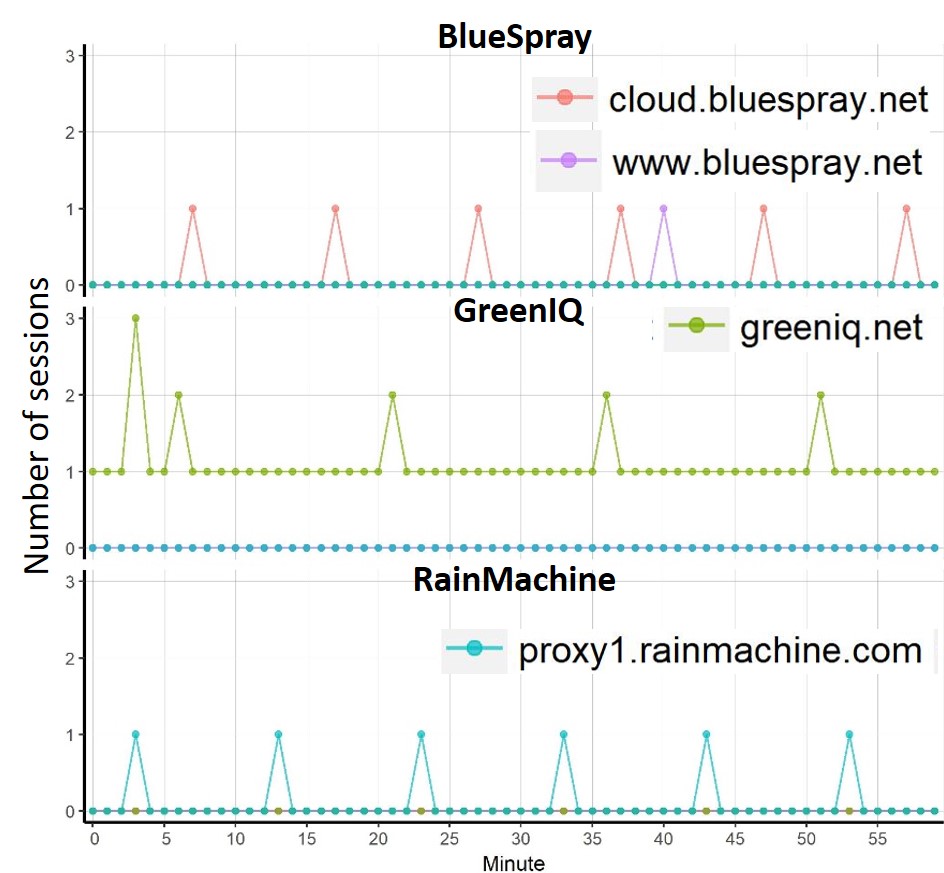}
\caption{Analysis of the number of TCP sessions opened by smart irrigation systems to their cloud servers during a typical hour.}
\label{fig:destinations-grpah}
\end{figure}

\begin{figure}
\centering
\includegraphics[width=0.8\columnwidth]{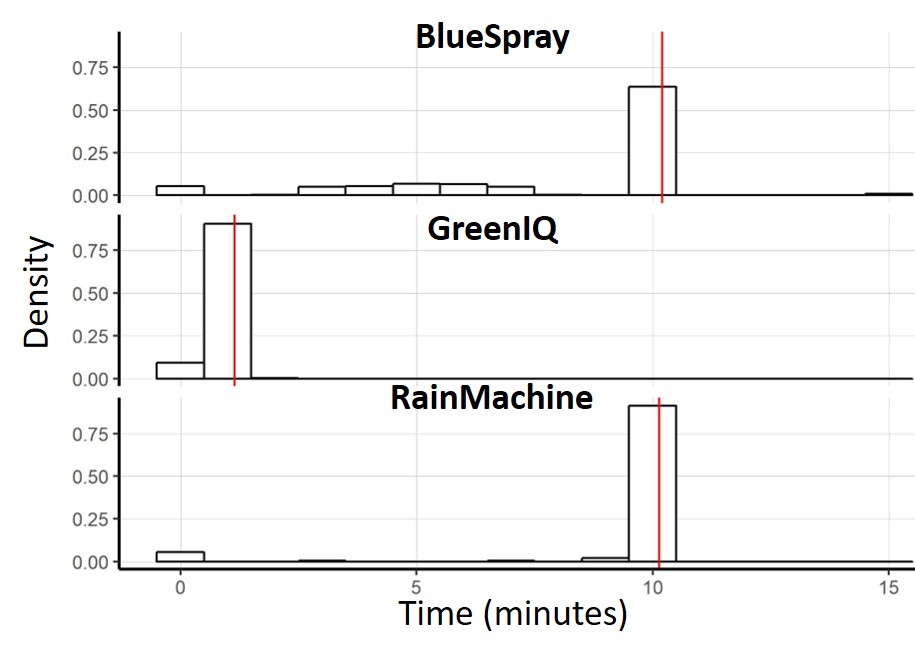}
\caption{Distribution of the time between two consecutive sessions. The red line represents the 99\% percentile for each model.
}
\label{fig:inter-arrival-time}
\end{figure}

As can be seen in Figure \ref{fig:destinations-grpah}, smart irrigation systems typically interact with their cloud servers several times per hour (6-11 times). We analyzed the distribution of the average time between two consecutive outgoing packets from any smart irrigation system to its cloud server. 
As can be seen in Figure \ref{fig:inter-arrival-time}, for the GreenIQ smart irrigation system, the average time between two consecutive sessions with its cloud server is much lower than that of BlueSpray and RainMachine. 
Overall, the maximum amount of time between two consecutive sessions with the cloud servers is 15 minutes (the 99\textsuperscript{th} percentile is approximately 10 minutes).

Based on this observation we present Algorithm \ref{alg:detection}, a smart irrigation system classification model. 

\begin{algorithm}
\caption{}\label{alg:detection}
\begin{algorithmic}[1]
\Procedure{isSmartIrrigationSystem}{ip,period}
\State	$bluespray1 = "cloud.bluespray.net"$
\State	$bluespray2 = "www.bluespray.net"$
\State	$greeniq = "www.greeniq.net"$
\State	$rainmachine = "proxy1.rainmachine.com"$
\State{startTime = currentTime()}
\State	$applyMitmAttackToTarget(ip)$
\For{$packet : nextPacket()$}	
	\State	$dstIP = packet.ip.dst$    
\If{dstIP == bluespray1}
\State $return \textbf{ BlueSpray}$
\EndIf
\If{dstIP == bluespray2}
\State $return \textbf{ BlueSpray}$
\EndIf
\If{dstIP == greeniq}
\State $return \textbf{ GreenIQ}$
\EndIf
\If{dstIP == rainmachine}
\State $return \textbf{ RainMachine}$
\EndIf
\If{startTime + period >= currentTime()}
\State $return \textbf{ None}$
\EndIf
\EndFor
\EndProcedure
\end{algorithmic}
\end{algorithm}

\begin{figure}
\centering
\includegraphics[width=0.8\columnwidth]{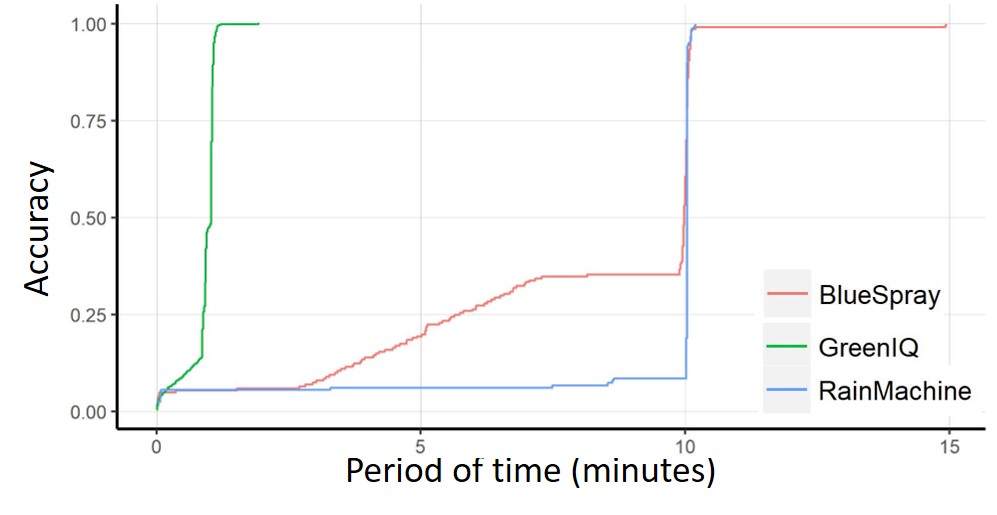}
\caption{Algorithm \ref{alg:detection}'s accuracy for various time periods.}
\label{fig:time-to-confidence}
\end{figure}

Algorithm \ref{alg:detection} receives as input an \textit{IP} of a suspicious device that is connected to the LAN of the bot and a \textit{period} of time for capturing traffic. It applies ARP spoofing to the suspicious \textit{IP} (line 7) and analyzes outgoing traffic from the \textit{IP} for the amount of time given by \textit{period}. It classifies the suspicious \textit{IP} as a smart irrigation system if the outgoing traffic is being sent to known cloud servers. If the period of time that was specified has passed, it classifies the suspicious \textit{IP} as \emph{other} device. Figure \ref{fig:time-to-confidence} presents accuracy results of applying Algorithm \ref{alg:detection} from a laptop connected to the same LAN as the smart irrigation systems for various periods of time. As can be seen in Figure \ref{fig:time-to-confidence}, the classification accuracy reaches 99.9\% after 10 minutes of analysis and 100\% after 15 minutes.

\section{Spoofing Attacks}
\label{section:MiTM}

In this section we present a set of \textbf{spoofing attacks} on commercial smart irrigation systems that a bot can implement (after detecting a connected smart irrigation system) in order to spoof the input of the irrigation system. We consider a spoofing attack that: (1) changes an input to a smart irrigation system, (2) can be applied remotely by the attacker from a bot running on a compromised device that is connected to the LAN, and (3) results in watering according to the attacker's wishes. Smart irrigation systems obtain information from cloud servers, weather forecast services, and sensors. All of the attacks presented in this section that were used to spoof smart irrigation system inputs are based on MITM attacks; the MITM attacks were applied by a bot running on another device that is connected to the same LAN and managed to intercept outgoing traffic sent from a smart irrigation system in order to impersonate to destination and hijack the entire session.



\subsection{Spoofing smart irrigation system configuration}


\begin{figure*}[]
  \centering
  \begin{minipage}[b]{0.6\textwidth}
    \includegraphics[width=\textwidth]{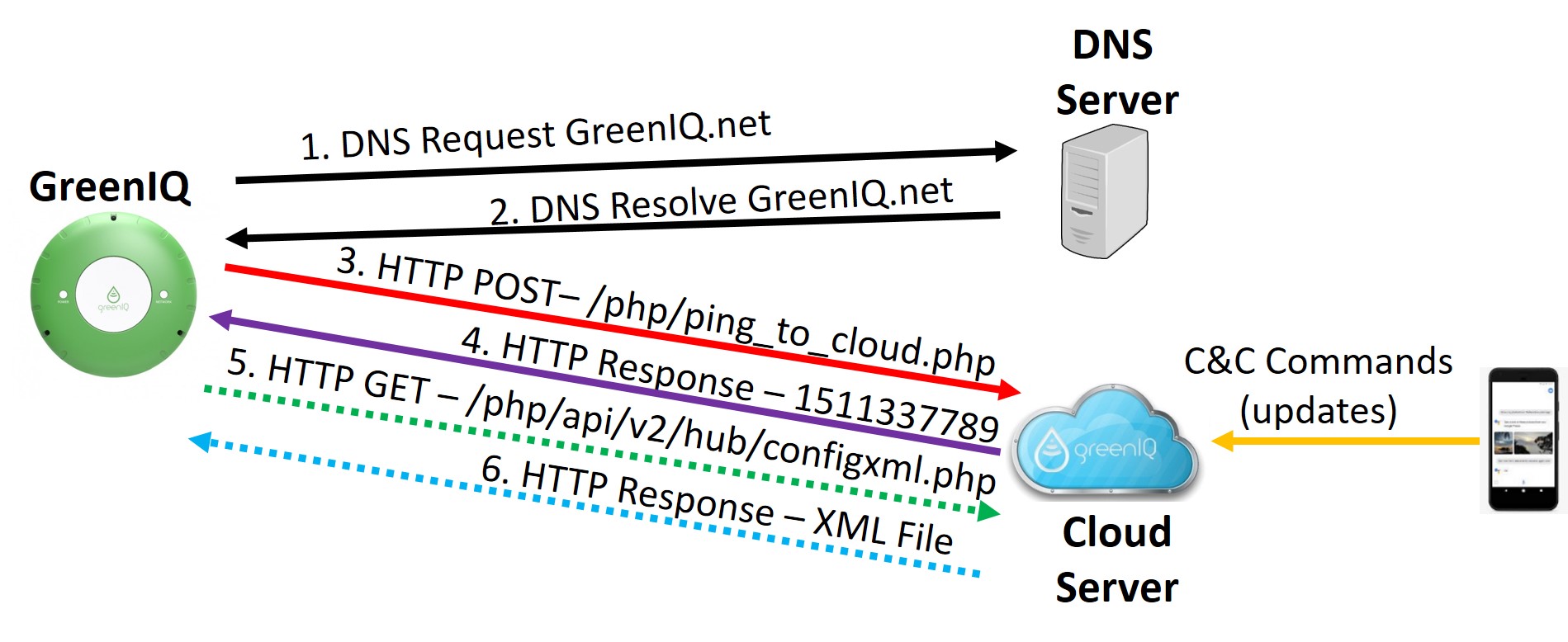}
    \caption{Session stages between the GreenIQ smart irrigation system and its cloud server.}
    \label{fig:cloud-interface}
  \end{minipage}
  \hfill     
  \begin{minipage}[b]{0.30\textwidth}
    \includegraphics[width=\textwidth]{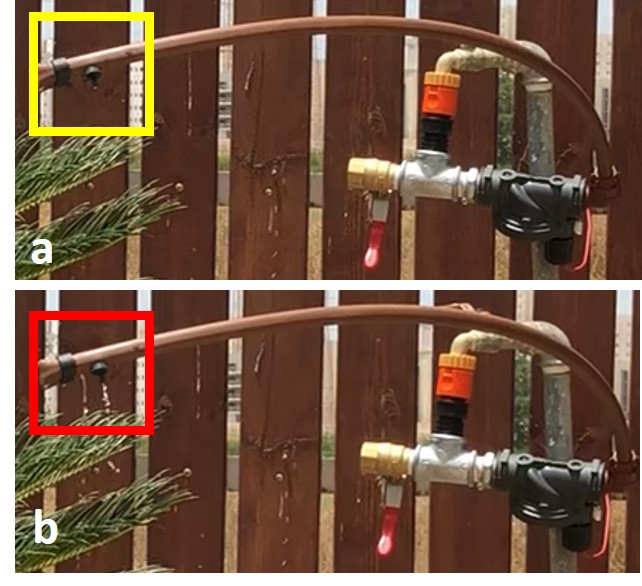}    
    \caption{A dry dropper boxed in yellow, and a dripping dropper (boxed in red) as a result of applying a watering plan injection attack.}
    \label{fig:greeniq-cloud}
  \end{minipage}
\end{figure*}

The attacks demonstrated in this subsection represent attempts to \textbf{spoof the smart irrigation system's configuration} response that is sent from the cloud server by impersonating the smart irrigation system's cloud server. We demonstrate this attack against the GreenIQ smart irrigation system.

\subsubsection{Vulnerability}

The cloud server is supposed to mediate between a C\&C device (e.g., smartphone application) which can be located anywhere around the world and a smart irrigation system. Figure \ref{fig:cloud-interface} outlines the interface between the GreenIQ application running on a smartphone to the GreenIQ smart irrigation system via the cloud server. Using a smartphone application, the user sends C\&C commands to the cloud server (yellow arrow in Figure \ref{fig:cloud-interface}). Independently, a \textbf{ping\_to\_cloud} request (that contains the user's ID) is launched from the GreenIQ smart irrigation system to the cloud server every minute in order to obtain the timestamp of the last time the user updated the watering plan configuration stored in the cloud server (red arrow in Figure \ref{fig:cloud-interface}). A response is sent from the cloud server with this timestamp (purple arrow in Figure \ref{fig:cloud-interface}). If the timestamp received from the cloud server is greater (after) than the timestamp that is stored on the GreenIQ smart irrigation system (signifying a more recent user update), a \textbf{configxml} request to retrieve the new watering plan configuration is launched by the GreenIQ smart irrigation system (green arrow in Figure \ref{fig:cloud-interface}). A response is sent from the cloud server with a file that contains the new watering plan configuration in XML format (blue arrow in Figure \ref{fig:cloud-interface}). This XML file contains details about all of the watering plans scheduled by the user (dates, hours, duration, zones/valves, etc.). Listing \ref{listing-cloud} presents the code that implements the abovementioned description which was extracted from the main.py file of GreenIQ's firmware.

\begin{lstlisting}[language=Python, breaklines= true, numbers	= left,numbersep=0pt,firstnumber = 312, showstringspaces=false,label = listing-cloud, xleftmargin=2em,framexleftmargin=1.5em,frame=single, captionpos=b,caption = GreenIQ's firmware code extracted from main.py file.]
# Check if config.xml was modified. If yes, retrieve it.
if new_config > current_config :
    main_log.info('config time updated. current_config: %d , new_config %d' % (current_config,new_config))
    s2 = GD.get_config_xml(hub_hash)
    if s2:
        current_config = new_config
        update_ping_to_cloud_immidiate = True
else:
    main_log.info('config time did not change. new_config: %d' % new_config)
\end{lstlisting}

As can be seen in Listing \ref{listing-cloud}, the timestamp configuration received from the cloud, $new\_config$, is being compared to $current\_config$, which is the timestamp stored in GreenIQ of the last time the user updated the watering plan configuration (line 313). If an update was made by the user, the new configuration is retrieved from the cloud server (line 315) and stored in GreenIQ (line 317).

\subsubsection{Exploitation}

We demonstrate how an attacker can (1) launch watering using the GreenIQ smart irrigation system by injecting his/her own watering plans, and (2) cause the GreenIQ smart irrigation system to deny service permanently, thereby preventing any remote C\&C interface with the smart irrigation system. Both attacks are applied by a bot that impersonates a weather forecast service. In our experiment we used the GreenIQ application to schedule a watering plan that waters 24/7 (every day, all day long) for a period of time between two future dates. We captured HTTP communication between the GreenIQ smart irrigation system and the cloud server during this time and extracted the watering plan configuration that was sent from the cloud server in the XML file. Then, using the GreenIQ application, we restored the GreenIQ smart irrigation system to its previous state. 

Algorithm \ref{alg:cloud} presents the exploitation code used to inject a watering plan for a given future time period.

\begin{algorithm}
\caption{}\label{alg:cloud}
\begin{algorithmic}[1]
\Procedure{SpoofConfiguration}{packet,start,end}
\State	$ping \gets "/php/ping\_to\_cloud.php"$
\State	$retrieve \gets "/php/api/v2/hub/configxml.php"$
	\State	$method \gets packet.http.request.method$   
    \State	$path \gets packet.http.request.uri.path$   

	\State	$dstIP \gets packet.ip.dst$    
\If{dstIP != "www.greeniq.net"}
\State $return$
\EndIf
\If{(method == "POST" \& path == ping)}
\State	$sendFakeTimestampResponse(end)$
\EndIf
\If{(method == "GET" \& path == retrieve)}
\State	$path \gets createFakeXML(start,end)$
\State	$sendFakeXMLResponse(path)$   
	\EndIf
\EndProcedure

\end{algorithmic}
\end{algorithm}

Algorithm \ref{alg:cloud} receives a $packet$ sent from the GreenIQ application and two future timestamps, $begin$ and $end$, to launch watering. First, it verifies that the $packet$ was sent to GreenIQ's cloud server (line 7). If the $packet$ is a $ping\_to\_cloud$ request, a fake timestamp (denoted by the received parameter $end$) is sent to the GreenIQ smart irrigation system by the bot (line 10). A response with a future timestamp will trigger another request to retrieve the updated XML configuration launched from the smart irrigation system. If the received packet is a $configxml$ request, a fake XML with a watering plan between the timestamps of $begin$ and $end$ is sent to the smart irrigation system by the bot (line 13).

We installed this code on a laptop that was connected to the same LAN as the GreenIQ smart irrigation system and applied ARP spoofing in order to refer traffic from the GreenIQ smart irrigation system to our bot. A $ping\_to\_cloud$ request is sent from the GreenIQ smart irrigation system to its cloud server every minute over HTTP communication; this request is intercepted by our code. 
Two snapshots demonstrating the attack are presented in Figure \ref{fig:greeniq-cloud}. As can be seen in the figure, the attack caused the GreenIQ smart irrigation system to launch watering immediately after the response was received from the bot. We consider this attack a \textbf{watering plan injection attack}. It allows the attacker to trigger the GreenIQ smart irrigation system (via the bot) to launch watering according to his/her wishes. In addition, since the GreenIQ smart irrigation system sends requests to its cloud sever every minute, a watering plan injection attack can be performed by the attacker close to the time of the DDoS attack making it harder for detection.


We analyzed the code that was extracted from the GreenIQ firmware, and this is presented in Listing \ref{listing-cloud}. As can be seen from line 312 (the if condition), this code verifies whether the received timestamp ($new\_config$) is greater (after) than the timestamp stored in the GreenIQ smart irrigation system ($current\_config$). If $new\_config$ is greater (after), the new timestamp is stored in the GreenIQ smart irrigation system, and the corresponding watering plan configuration is retrieved from the cloud server. No other verification regarding the correctness of the timestamp received, stored in $new\_config$ is performed. This can be exploited by the attacker who can use the bot in order to cause the GreenIQ smart irrigation system to \textbf{permanently deny service} by implementing Algorithm \ref{alg:cloud} with an $end$ timestamp value that is far into the future (e.g., the timestamp of 1/1/2022). By applying the following, the bot causes the GreenIQ smart irrigation system to ignore any C\&C command that is launched by the user until the time that is mentioned in the response, because any C\&C command during this period of time will not be considered by the GreenIQ smart irrigation system as a user update (line 313 of the code in Listing \ref{listing-cloud}). By combining a \textbf{permanent denial of service attack} (by replying with a future watering plan, e.g., the timestamp of 1/1/2022), with a \textbf{watering plan injection attack} that triggers the GreenIQ smart irrigation system to launch watering 24/7, the bot causes the irrigation system to start watering indefinitely and prevents the user from remotely stopping the watering using a C\&C device. The only way in which the GreenIQ owner can stop the GreenIQ smart irrigation system from watering in this attack scenario is by physically turning off the main water line. In order to restore the GreenIQ smart irrigation system regular operation, the user would have to apply a factory reset to delete the future timestamp.

\subsection{Spoofing weather forecast}

The attacks demonstrated in this subsection represent attempts to \textbf{spoof the weather forecast} response sent from a weather forecast server by impersonating a weather forecast service. We demonstrate this attack against the RainMachine smart irrigation system.

\subsubsection{Vulnerability}
The RainMachine smart irrigation system was designed to save water and money by automatically adapting its watering plan to weather forecasts. It allows the user to configure a base watering plan according to the amount of water that is needed to water his/her yard and plants. Given the base watering plan configuration and the weather forecast (obtained from weather forecast services), the RainMachine smart irrigation system adapts its watering plan automatically. This means that for a rainy/cold weather forecast, watering will not take place, or only a percentage of the amount of water required by the base watering plan will be used (just the amount needed in order to fulfill the water requirements specified in the user's configuration). In cases in which there is a forecast for dry weather, the RainMachine smart irrigation system automatically adjusts itself to compensate for a lack of precipitation by supplementing with watering plans that consume the required amount of water, based on the user's configuration of the base watering plan.
We analyzed the RainMachine smart irrigation system's firmware and found the \textit{MainActivity.java} file. RainMachine smart irrigation system contains a touchscreen that presents the weather forecast for the upcoming week. In addition, it presents the percentage of water that the smart irrigation system plans to consume in order to fulfill the water requirements specified in the base watering plan configured by the user. We searched for the code that calculates the exact percentage of water that is going to be consumed by the RainMachine smart irrigation system each day during the upcoming week and found that it relies on the amount of rain that is forecast for each day, as can be seen in Listing \ref{listing-mainactivity}.

\begin{lstlisting}[language=Java, breaklines= true, numbers = left,numbersep=0pt,firstnumber = 370, showstringspaces=false,label = listing-mainactivity, xleftmargin=2em,framexleftmargin=1.5em,frame=single,captionpos=b, caption = RainMachine's firmware code from MainActivity.java file]
int percentValue = Math.round(100.0f * ((Float) ((MainDayViewModel) viewModel.days.get(startDate.plusDays(indexDay))).programWaterNeed.get(viewModel.indexProgram)).floatValue());
\end{lstlisting}

\begin{figure*}[]
  \centering
  \begin{minipage}[b]{0.60\textwidth}
    \includegraphics[width=\textwidth]{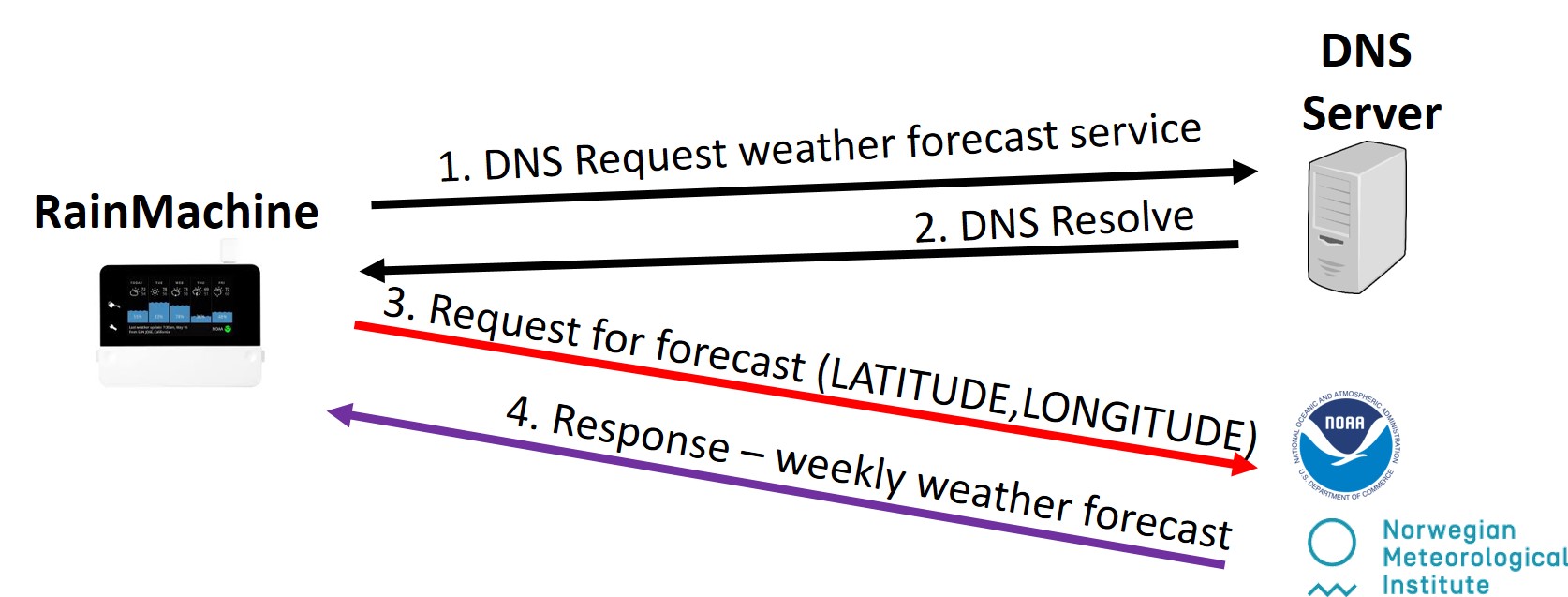}
    \caption{Session stages between RainMachine smart irrigation system and Met.no weather forecast service}
    \label{fig:weather-forecast-interface-1}
  \end{minipage}
  \hfill     
  \begin{minipage}[b]{0.30\textwidth}
    \includegraphics[width=\textwidth]{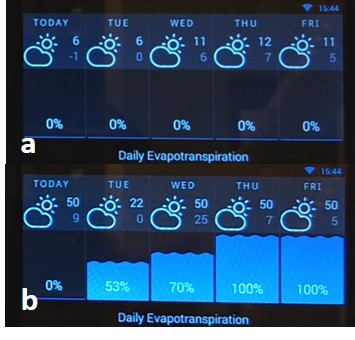}    
    \caption{The original weather forecast in London (upper picture) was spoofed to a fake weather forecast (lower picture)}
    \label{fig:weather-forecast-interface-2}
    
  \end{minipage}
\end{figure*}

We continued to analyze the RainMachine smart irrigation system's firmware searching for the word "\textit{Weather}." Listing \ref{weather-services} presents code from the $ParserResponse.java$ file of weather forecast services that the RainMachine smart irrigation system interfaces with.

\begin{lstlisting}[language=Java, breaklines= true,numbersep=0pt,
linerange={1-1,3-3,5-5,7-7,9-9,11-11,13-13,15-15,17-17}, showstringspaces=false,label = weather-services, xleftmargin=2em,framexleftmargin=1.5em,frame=single,captionpos=b, caption = List of weather services extracted from RainMachine firmware ]
    public boolean isNOAA() 
    public boolean isMETNO() 
    public boolean isWUnderground() 
    public boolean isForecastIO() 
    public boolean isNETATMO() 
    public boolean isCIMIS() 
    public boolean isFAWN() 
    public boolean isWeatherRules() 
    public boolean isPWS() 
\end{lstlisting}

We searched for these names on the Internet and found the weather forecast services that appear in Listing \ref{weather-services}. We analyzed the REST API for each weather forecast service that was found in Listing \ref{weather-services}. We note the following interesting observation: during the time in which this research was conducted, most of the weather forecast services provided a REST API based on HTTP communication. Figure \ref{fig:weather-forecast-interface-1} presents the REST API interface between the RainMachine smart irrigation system and a weather forecast service. An HTTP request that contains RainMachine's location (in latitude-longitude format) is sent from the RainMachine smart irrigation system to a weather forecast service. A response is sent from the weather forecast service in the form of a file in XML format that contains the weather forecast (hourly resolution) with various details including: temperature, wind direction and speed, cloudiness, humidity, barometric pressure, etc. Four requests per day are launched by the RainMachine smart irrigation system to the weather forecast service, and based on the weather forecast received, the RainMachine smart irrigation system adjusts its future watering plans.

\subsubsection{Exploitation}
We demonstrate how an attacker can manipulate the RainMachine smart irrigation system to schedule unnecessary watering plans based on his/her wishes by impersonating a weather forecast service and injecting a fake weather forecast. We analyzed the Met.no API and found that it provides a REST interface based on HTTP communication. We identified the format of the response sent from the Met.no weather forecast service, and based on these findings, we wrote a Python code that changes weather forecast parameters between two given timestamps. 



We installed our code on a laptop that was connected to the same LAN as the RainMachine smart irrigation system and implemented an ARP spoofing attack to refer traffic sent from the RainMachine smart irrigation system to the Met.no weather forecast service. Originally, the RainMachine smart irrigation system was configured to work in London. We performed the attack during the winter; since London is rainy in the winter, no watering would likely be needed in order to fulfill the requirements of the base watering plan configuration. Accordingly, RainMachine adapted its watering plan to consume no water for the upcoming week, as can be seen in Figure \ref{fig:weather-forecast-interface-2}a which presents RainMachine smart irrigation system's screen before the attack.

A request to the Met.no weather forecast service is sent every six hours from the RainMachine smart irrigation system over HTTP communication and in this attack such a request was intercepted by our code. 
As can be seen in the Figure \ref{fig:weather-forecast-interface-2}a, the original weather forecast for London did not require any watering at all, because the temperatures forecasted were between -1\degree and 12\degree for the entire week. However, implementing the attack caused this temperature to be changed to values between 0\degree and 50\degree. As a result, the RainMachine smart irrigation system immediately adjusted its watering plan to compensate for these temperatures by scheduling watering plans, as can be seen in Figure \ref{fig:weather-forecast-interface-2}b.

Another way of manipulating the RainMachine smart irrigation system in order to schedule unnecessary watering plans is by changing the location of the request sent from the RainMachine system to the weather forecast service to the most arid place on Earth for the day on which the attack is performed. The previous attack, which responds with a fake XML file, required the attacker to identify the format of the XML response that is sent from the weather forecast service. The current attack requires a much simpler process of changing the request location (longitude, latitude) value that is supplied as part of the request. We conducted an additional experiment in which we performed the attack by changing the location of the request from London to Algeria which was the driest and hottest place on Earth when we performed the attack. As a result of the attack, the RainMachine smart irrigation system adapted itself automatically to compensate for the dry weather and scheduled watering plans. Although the response sent from the Met.no weather forecast service contained the coordinates of the city in Algeria, the RainMachine smart irrigation system did not identified this change and accepted the new weather forecast. As a result, it adjusted its watering plan to compensate for the lack of water and the weather in Algeria.

We consider these attacks \textbf{weather forecast injection attacks}. They allow an attacker to trigger the RainMachine smart irrigation system (via the bot) to launch watering based on his/her wishes. In addition, since the RainMachine smart irrigation system sends requests to weather forecast services every six hours, a weather forecast injection attack can be performed by the attacker around the time of the DDoS attack.

\subsection{Sensor attacks}

Many smart irrigation systems allow sensor connectivity, using sensors like rain sensors, water flow sensors, and soil moisture sensors to regulate watering and water consumption more efficiently. IoT device sensor attacks are very common and can appear in one of the following ways:

\begin{itemize}
\item \textbf{Compromising a sensor} - the attacker manages to compromise a sensor (e.g., using a supply chain attack, exploiting an OS vulnerability). As a result, the sensor sends false data.

\item \textbf{Spoofing outgoing communication from the sensor} - the attacker manages to change the data that is sent from the sensor (e.g., using a MITM attack).

\item \textbf{Physically influencing the sensor} - the attacker manages to influence the phenomena that is being measured (e.g., by hitting a temperature meter, pouring water on a rain or water moisture sensor) so false measurements were obtained.
\end{itemize}

Spoofing a sensor's output with any of the abovementioned methods will influence the operation of smart irrigation systems. Smart irrigation systems with a connected rain sensor allow the user to define rules that prevent watering on rainy days. Considering this fact, spoofing a rain sensor's data so that it won't notify the smart irrigation system when it is rainy causes the daily watering program to work as usual instead of being disabled during rainy weather.

However, since smart irrigation systems were just introduced a few years ago, the current generation of commercial smart irrigation systems supports only wire connectivity to a reserved set of connectors that can be found on the smart irrigation system SoC board. This fact limits the type of attacks that can be performed by the attacker in order to spoof sensor data because: (1) spoofing outgoing communication from the sensors requires physical access to the cable that connects the sensor with the smart irrigation system, (2) physical attacks (the third type of attack mentioned above) on many sensors are not practical, since they require many people to engage with the sensors of the attacked smart irrigation system during the time of the attack, and (3) compromising a massive amount of sensors can only be done using a supply chain attack, which is not easy to perform.

We believe that next generation of smart irrigation systems will support wireless connectivity to sensors, creating a new attack vector for spoofing data that is sent from the sensor remotely to the smart irrigation system.

\section{Replay Attacks}
\label{section:replay-attacks}

In this section we present a set of \textbf{replay attacks} that can be implemented by a bot against a commercial smart irrigation system in order to launch watering. A replay attack (or playback attack) is a form of network attack in which a valid data transmission is maliciously or fraudulently transmitted. We consider a replay attack that: (1) can be applied remotely by the attacker by a bot running on a compromised device that is connected to the LAN, (2) results in watering according to attacker's wishes, and (3) exploits a legitimate HMI interface for C\&C as a means of attack. The attacks demonstrated in this section were performed by a bot that is running on another device that is connected to the same LAN as the smart irrigation system and generates 


\subsection{Scheduling a watering plan}

The attack demonstrated in this subsection is scheduling watering plan attack. We demonstrate this attack against the BlueSpray smart irrigation system.

\subsubsection{Vulnerability}
All smart irrigation systems provide an HMI to a C\&C device. The HMI can be operated from various C\&C devices including a mobile application, Web browser, or smart assistant. Using a C\&C device, the user can use the HMI to: (1) connect the smart irrigation system to a LAN, (2) update the watering plan configuration, (3) monitor the watering history, (4) define zones, (5), add sensors, etc. BlueSpray provides an HMI interface based on PCs and laptops via a Web browser that is based on HTTP communication. The user can open a Web browser (Chrome, Firefox, etc.) from another device that is connected to the same LAN, type BlueSpray's IP address, and send it C\&C commands. Listing \ref{bluespray-watering} presents a payload (JSON format) extracted from an HTTP packet for scheduling a watering plan that was sent from a Chrome browser to the BlueSpray smart irrigation system.

\begin{lstlisting}[language=XML, breaklines= true, numbers	= left,numbersep=0pt, showstringspaces=false,label = bluespray-watering, xleftmargin=2em,framexleftmargin=1.5em,frame=single, captionpos=b,caption = Payload of an HTTP request sent to BlueSpray]
{"action":"set","data":[{"enabled":1,"type":2,"program":10,"rpt":[0],"season":0,"cycle":[5,60],"name":"New run","start_date":"2018-06-17","start_time":0,"id":5,"flag":"change"}],"msgid":77080}
\end{lstlisting}

We were surprised to find that no authentication is required in order to communicate with the BlueSpray smart irrigation system from another device that is connected to the same LAN.
\begin{figure}
\centering
\includegraphics[width=0.9\columnwidth]{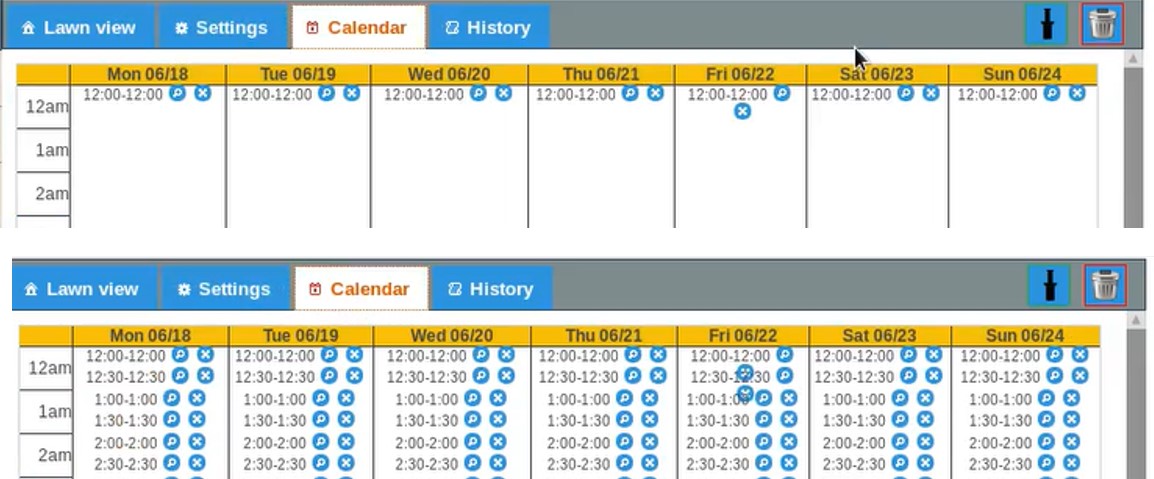}
\caption{BlueSpray's Web user interface. Before the attack (upper picture) there are no watering plans, and after the attack (lower picture) watering plans have been scheduled for the entire week.}
\label{fig:bluespray}
\end{figure}
\subsubsection{Exploitation}
We demonstrate how an attacker can launch watering via the BlueSpray smart irrigation system by scheduling watering plans according to his/her wishes. We analyzed the HTTP packets of watering plan updates sent from a laptop to the BlueSpray smart irrigation system from a PC connected to the same LAN via the Chrome Web browser and learned how such a request is generated. Based on our findings, we wrote a Python code that schedules watering between two given timestamps using HTTP request that is sent to the BlueSpray smart irrigation system. 

We reset the BlueSpray smart irrigation system to its previous configuration with no watering plans. We installed our code on a laptop that was connected to the same LAN and ran the code. The code launched an HTTP request to schedule watering plans for the entire week. 
Two snapshots that demonstrate this attack scenario are presented in Figure \ref{fig:bluespray}.
As can be seen in Figure \ref{fig:bluespray}, our code successfully scheduled new watering plans for the BlueSpray smart irrigation system.


\subsection{Opening the valves of a smart irrigation system}

The attacks demonstrated in this subsection were implemented on the GreenIQ smart irrigation system. 

\subsubsection{Vulnerability}
We analyzed the GreenIQ smart irrigation system's firmware and looked for the code that opens a valve. Listing \ref{set-gpio} presents code from the $greeniq\_defs.py$ file.

\begin{lstlisting}[language=Python, breaklines= true, numbers = left,numbersep=0pt,firstnumber = 221, showstringspaces=false,label = set-gpio, xleftmargin=2em,framexleftmargin=1.5em,frame=single, caption = A code for opening a valve (extracted from GreenIQ firmware)]
def set_gpio(MAX_PORTS, gpio_map, gpio_command, high_is):
    global model_utilities
    model_utilities.set_gpio(MAX_PORTS, gpio_map, gpio_command, high_is)
\end{lstlisting}

A GPIO (general purpose input output) interface is used by the GreenIQ smart irrigation system's SoC board (Raspberry Pi) to control the connected valves. We looked in the firmware's code for a specific call to the function $set\_gpio()$ and found the following code (presented in Listing \ref{gpio}) in the $greeniq\_defs.py$ file:

\begin{lstlisting}[language=Python, breaklines= true, numbers = left,numbersep=0pt,firstnumber = 427, showstringspaces=false,label = gpio, xleftmargin=2em,framexleftmargin=1.5em,frame=single, caption = Execution of the code for opening a valve (Listing \ref{gpio})]
# Testing - Operate Master Valve
set_gpio(MAX_PORTS, gpio_map, '00000010', high_is)
time.sleep(test_preiod)
\end{lstlisting}
Closing the valves is handled by executing the following code: $set\_gpio (MAX\_PORTS, gpio\_map, '00000000', high\_is)$. 

\subsubsection{Exploitation}
\begin{figure}
\centering
\includegraphics[width=1.0\columnwidth]{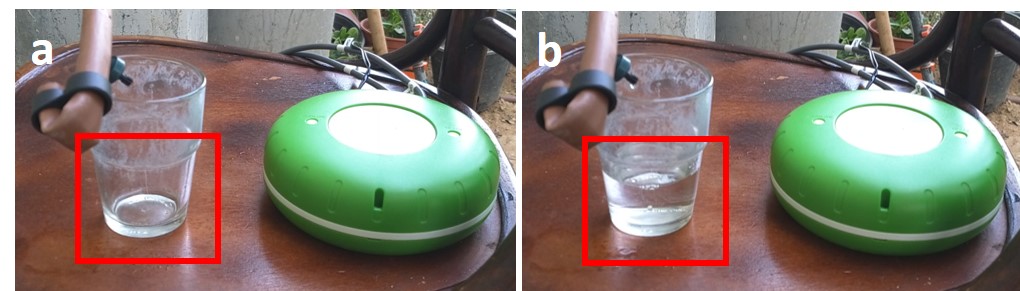}
\caption{(a) an empty glass, and (b) a glass that is being filled as a result of a compromised device that uses SSH communication to launch watering}
\label{fig:greeniq-ssh}
\end{figure}

We demonstrate an opening valve attack by opening and closing the master valve every 10 seconds using SSH communication from a laptop that is connected to GreenIQ smart irrigation system's LAN. 
Two snapshots from the experiment are presented in Figure \ref{fig:greeniq-ssh}. 

As can be seen in Figure \ref{fig:greeniq-ssh}, watering starts and ends every 10 seconds. An \textbf{opening valve attack} can be implemented from a bot running on: (1) a compromised device connected to the LAN of the GreenIQ smart irrigation system using SSH communication and a password (as we did), and (2) the GreenIQ smart irrigation system itself. This allows the attacker to trigger the GreenIQ smart irrigation system (via the bot) to launch watering according to his/her wishes.

\section{Calculating the Damage}
\label{section:damage}

\begin{figure*}
\centering
\includegraphics[width=0.65\textwidth]{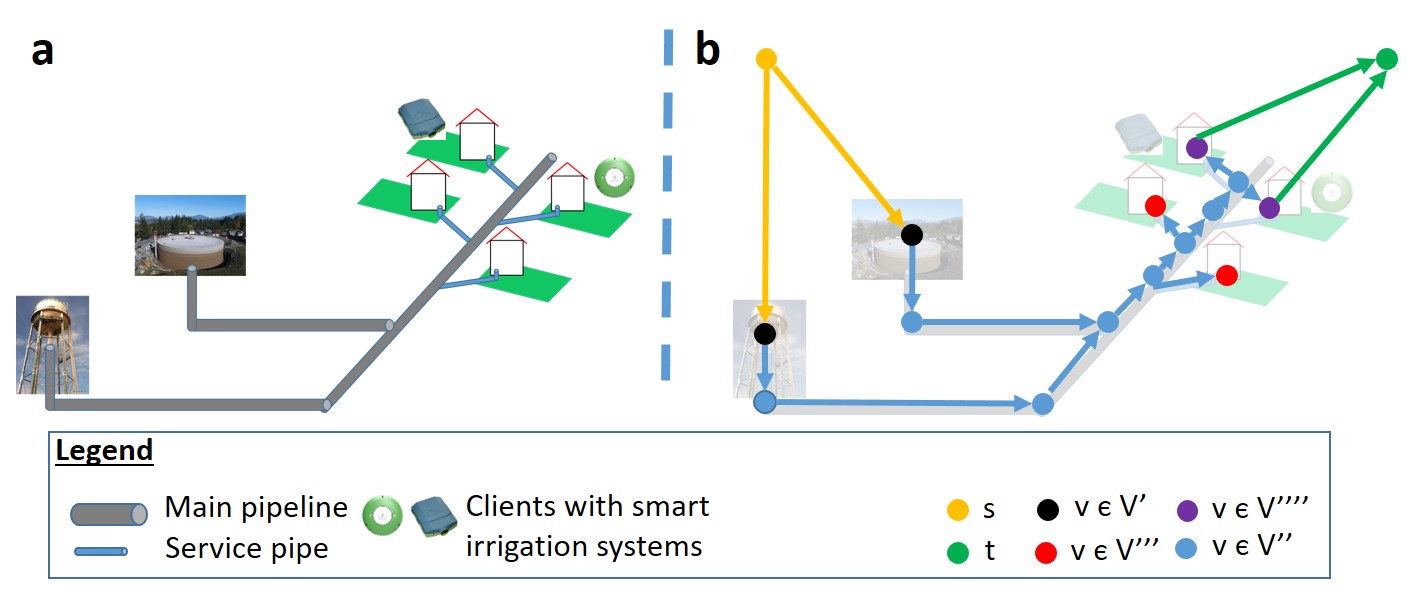}
\caption{Modeling pipeline system that distributes water as flow network}
\label{fig:flow-network}
\end{figure*} 

In this section we describe two methods to calculate the damage that can be caused by applying a distributed attack on urban water services: (1) a theoretical calculation using a flow network, and (2) an empirical estimation using an experiment.

\subsection{Calculating the damage using a flow network}
Given an area with a pipeline system that distributes water obtained from an areal water reservoir to clients/sinks (e.g., homes, yards/gardens, public locations, etc.), we calculate potential water waste and financial damage by modeling the area's pipeline system as a flow network and identifying maximum flow in the network with well-known algorithms. A flow network is defined as quartet of (G,c,s,t), where \textit{G = (V,E)} is a directed graph, \textit{c} is a capacity function, \textit{s} is a source vertex, and \textit{t} is a target vertex. Given an areal distribution pipeline with water providers (e.g., water reservoirs, water tanks), water consumers (e.g., houses, schools, etc.), and a network of pipelines, we build a flow network as follows:

\textbf{Vertices} 
\begin{enumerate}
\item Let V' be a set of vertices, where v belongs to V' if it is a water reservoir. 
\item Let V'' be a set of vertices, where v belongs to V'' if it is a pipeline junction.
\item Let V''' be a set of vertices, where v belongs to V''' if it is a sink/water consumer (e.g., house, school, etc.), and a smart irrigation system is not connected to the consumer.
\item Let V'''' be a set of vertices, where v belongs to V'''' if it is a sink/water consumer (e.g., house, school, etc.), and a smart irrigation system is connected to the consumer.
\item Let us define a new supersource vertex \textbf{s} and new supersink vertex \textbf{t}.
\end{enumerate}

The entire set of vertices \textbf{V} is defined as follows: 
$V = V' \cup V'' \cup V''' \cup V'''' \cup \{s,t\}$

\textbf{Edges} 
\begin{enumerate}
\item Let E' be a set of edges, where e = (v1,v2) belongs to E' if a pipeline between v1 and v2 exists (v1,v2 belongs to V).
\item Let E'' be a set of edges between the source vertex s to every vertex in V'. 
\item Let E''' be a set of edges between every sink vertex v (belongs to V'''') that has a smart irrigation system to the target vertex t. 

\end{enumerate}
The entire set of edges \textbf{E} is defined as follows: $E = R' \cup E'' \cup E'''$
The entire graph G is defined by \textbf{G = (V,E)}

\textbf{Capacity function}

Let c be a capacity function from edges to a real number defined as follows:
 \begin{equation}
    c (v1,v2) =
    \begin{cases*}
      \infty & if  $v1 \in V'$ \\
      w & if  $v2 \in V''$ \\
      \begin{tabular}[c]{@{}l@{}}			
corresponding\\
pipeline\\
capacity
 \end{tabular} & otherwise
    \end{cases*}    
  \end{equation}

where \textit{w} is the average amount of water that can be consumed by a smart irrigation system.

The flow network is defined by (G,c,s,t). An edge from a vertex to \textit{t} in a flow network (G,c,s,t) exists only if the vertex contains a connected smart irrigation system. Let us assume that each bot can consume the maximum amount of water that can be supplied to the household by opening all of the smart irrigation valves. The potential water wasted can be calculated by applying well- known algorithms for maximum flow problems (Ford-Fulkerson, Edmonds-Karp, MPM, and Dinic's algorithm) on the constructed flow network and determining the amount of water flow that is found in the supersink vertex \textit{t}. The \textbf{financial damage} can be calculated by multiplying the following: the maximum flow of the supersink vertex \textit{f}, the average combined water tariff in the area, and the time period of the attack.

\subsection{Estimating the damage using an experiment}

An alternative way to estimate the damage that can be caused by applying a distributed attack on urban water services is by calculating the amount of water that can be consumed from a sprinkler. A typical sprinkler's water flow is between 0.66 to 4.93 cubic meters per hour (as can be seen in the specs of the Falcon 6504 sprinkler \cite{Falcon}). Let us assume that the attacker controls a botnet of $N$ smart irrigation systems (each of which is connected to a single sprinkler) which are operated for a given period of time $t$. The expected water waste caused by applying the attack is calculated by multiplying the average water flow (2.795 cubic meters per hour) by the size of the botnet and the amount of time (the duration of the attack): 

\begin{equation}
  \textit{wasted water ($m^3$)} = 2.795 \times N \times T
  \label{eq:lagrange}
\end{equation} 

Table \ref{tab:amount-waste} presents the calculation of the damaged that can be caused (wasted water) by applying the attack with various numbers of bots and periods of time.


\begin{table}[tbp]
    \centering
    \caption{Damage Calculation}~\label{tab:amount-waste}
	\resizebox{0.8\columnwidth}{!}{%
    \begin{tabular}{|p{5.6em}|p{4.2em}|p{4.95em}|p{5em}|}    
\cmidrule{1-3}    \multicolumn{1}{|p{5.6em}|}{\textbf{Botnet size (number of sprinklers)}} & \textbf{Amount of time} & \textbf{Average amount of water wasted} & \multicolumn{1}{r}{} \\
\cmidrule{1-3}    1     & 1 hour & 2.795 $m^3$ & \multicolumn{1}{r}{} \\
    \midrule
    1,355 & 1 hour & \multirow{2}[4]{*}{3,787 $m^3$ } & \multirow{2}[4]{*}{\begin{tabular}[c]{@{}l@{}} Typical\\water tower\\capacity
\end{tabular}} \\
\cmidrule{1-2}    13,550 & 6 minutes & \multicolumn{1}{c|}{} & \multicolumn{1}{c|}{} \\
\midrule
    143,200 & 1 hour & \multirow{2}[4]{*}{404,244 $m^3$ } & \multirow{2}[4]{*}{\begin{tabular}[c]{@{}l@{}} Floodwater\\reservoir\\capacity
\end{tabular}} \\
\cmidrule{1-2}    23,866 & 6 hours & \multicolumn{1}{c|}{} & \multicolumn{1}{c|}{} \\
    \bottomrule
    \end{tabular}%
	}
\end{table}




A standard water tower capacity is 3,785 cubic meters (according to \cite{water-tower}) and as can be seen in Table \ref{tab:amount-waste}, it requires a botnet of 1,355 sprinklers that water a single hour in order to waste 3,787 cubic meters, a volume of water which is greater than capacity of a standard water tower. A small floodwater reservoir capacity is 400,000 cubic meters (e.g., Betarim \cite{betarim}) and as can be seen in Table \ref{tab:amount-waste}, it requires a botnet of 23,866 sprinklers that water six hours (overnight) in order to waste 404,244 cubic meters, a volume of water which is greater than capacity of a small floodwater reservoir.

\section{Countermeasures}
\label{section:countermeasures}
In this section we describe countermeasures to detect and prevent a distributed attack against urban water services. A distributed attack launched from smart irrigation systems can be detected by deploying a model that monitors unusual water consumption in urban water services (e.g., using anomaly detection methods). However, even if such an attack can be detected by an urban watering service, its ability to react to such an attack is very limited. The only thing that an urban watering service can do when such an attack is detected is stop water distribution. While this solution prevents the attacker from wasting any more water, it also prevents people from obtaining water which is the aim of the attacker. Preventing people from obtaining a resource from critical infrastructure can even be considered a national disaster, as was the case in the cyber attack against the Ukrainian power grid  \cite{cyberattack-on-Ukrainian-power-grid}. Preventing a bot from impersonating a party that a smart irrigation system interfaces with can be done by upgrading HTTP communication to HTTPS communication. Doing this will prevent the attacker from spoofing TCP packets . In addition, SSH communication is not needed in order to communicate with a smart irrigation system when a cloud serves as a mediator, so disabling SSH communication will prevent attackers from executing a code on smart irrigation systems by detecting weak passwords.

\section{Ethical Considerations and Disclosure}

We performed full ethical disclosure, revealing the vulnerabilities discussed in this paper and providing all of the relevant technical details and suggestions for addressing them to GreenIQ, RainMachine, and BlueSpray in June 2018. We received confirmation of our findings from each of them. GreenIQ thanked us for sharing our findings and decided to apply HTTPS communication between their smart irrigation system and cloud server. In addition, they decided to close the SSH port in their firmware to prevent an attacker from running Python code for watering. In June 2018 the Norwegian Meteorological Institute (Met.no) upgraded their HTTP API to an HTTPS version.

\section{Discussion}
\label{section:discussion}
The distributed attack described in this paper can result in (1) the DOS of water service in cities in which water is not provided by a natural water source (e.g., groundwater), and (2) financial damage. The proposed IoT botnet can also be used to attack other types of critical infrastructure as well. For example, it can be used to attack the smart grid which uses smart homes to produce electricity in order to implement a DoS attack on power distribution services (another critical infrastructure) in a neighborhood, as opposed to performing an attack directly on the regional electricity company. Another interesting method for triggering the attack, that does not require to compromise a device that is connected to a LAN of smart irrigation system, targets smart assistants which can be used to control smart irrigation systems. For example, an attacker could replicate Burger King's method and launch a Google query to Google Home (a smart assistant) by placing an ad on national television that contains an embedded message which initiates watering \cite{burger-king-against-google-home}. The Google query could even be launched via ultrasound \cite{DBLP:journals/corr/abs-1708-09537} using the advertisement. Given that recently malware has been used to attack smart refrigerators, air conditioning systems, thermostats, TVs, and now smart irrigation systems, we can only hypothesize whether the next generation of technicians will have to become cyber security analysts. The question remains: will Wireshark replace the traditional monkey wrench and Phillips screwdriver?

\bibliographystyle{IEEEtran}
\bibliography{IEEEabrv,main}

\begin{thebibliography}{10}
\providecommand{\url}[1]{#1}
\csname url@samestyle\endcsname
\providecommand{\newblock}{\relax}
\providecommand{\bibinfo}[2]{#2}
\providecommand{\BIBentrySTDinterwordspacing}{\spaceskip=0pt\relax}
\providecommand{\BIBentryALTinterwordstretchfactor}{4}
\providecommand{\BIBentryALTinterwordspacing}{\spaceskip=\fontdimen2\font plus
\BIBentryALTinterwordstretchfactor\fontdimen3\font minus
  \fontdimen4\font\relax}
\providecommand{\BIBforeignlanguage}[2]{{%
\expandafter\ifx\csname l@#1\endcsname\relax
\typeout{** WARNING: IEEEtran.bst: No hyphenation pattern has been}%
\typeout{** loaded for the language `#1'. Using the pattern for}%
\typeout{** the default language instead.}%
\else
\language=\csname l@#1\endcsname
\fi
#2}}
\providecommand{\BIBdecl}{\relax}
\BIBdecl

\bibitem{Top-10-smart-cities-1}
\BIBentryALTinterwordspacing
businessinsider, ``These 10 cities are the most prepared for the future,''
  2017. [Online]. Available:
  \url{http://www.businessinsider.com/smart-cities-ranking-easypark-group-2017-11/#10-melbourne-australia-received-a-perfect-score-on-its-4g-connectivity-1}
\BIBentrySTDinterwordspacing

\bibitem{smart-africa}
\BIBentryALTinterwordspacing
smartafrica, ``2017 transform africa to focus on smart cities.'' [Online].
  Available:
  \url{https://smartafrica.org/?Imijyi-ya-Afurika-igiye-kwigira-kuri-Kigali-muri-gahunda-ya-Smart-Africa}
\BIBentrySTDinterwordspacing

\bibitem{cyberattack-on-Ukrainian-power-grid}
\BIBentryALTinterwordspacing
T.~Hill, ``Russia tied to cyberattack on ukrainian power grid,'' 2016.
  [Online]. Available:
  \url{http://thehill.com/policy/cybersecurity/264794-russia-tied-to-cyberattack-on-ukrainian-power-grid}
\BIBentrySTDinterwordspacing

\bibitem{barcelona-irrigation-1}
M.~T. Review, ``Barcelona's smart city ecosystem,''
  \url{https://www.technologyreview.com/s/532511/barcelonas-smart-city-ecosystem/}.

\bibitem{RainMachine}
\BIBentryALTinterwordspacing
``Rainmachine - forecast smart wifi irrigation controllers.'' [Online].
  Available: \url{http://www.rainmachine.com/}
\BIBentrySTDinterwordspacing

\bibitem{BlueSpray}
\BIBentryALTinterwordspacing
``Bluespray - web based, wireless (wifi) irrigation controller.'' [Online].
  Available: \url{https://www.bluespray.net/}
\BIBentrySTDinterwordspacing

\bibitem{GreenIQ}
\BIBentryALTinterwordspacing
``Green iq - start saving water with greeniq.'' [Online]. Available:
  \url{https://greeniq.com/}
\BIBentrySTDinterwordspacing

\bibitem{Critical-infrastructure-EU}
\BIBentryALTinterwordspacing
E.~Commission, ``Critical infrastructure,'' 2017. [Online]. Available:
  \url{https://ec.europa.eu/home-affairs/what-we-do/policies/crisis-and-terrorism/critical-infrastructure_en}
\BIBentrySTDinterwordspacing

\bibitem{DHS-Critical-Infrastructure-Sectors}
\BIBentryALTinterwordspacing
D.~of~Homeland~Security, ``Critical infrastructure sectors,'' 2017. [Online].
  Available: \url{https://www.dhs.gov/critical-infrastructure-sectors}
\BIBentrySTDinterwordspacing

\bibitem{5560697}
S.~Shin, T.~Kwon, G.~Y. Jo, Y.~Park, and H.~Rhy, ``An experimental study of
  hierarchical intrusion detection for wireless industrial sensor networks,''
  \emph{IEEE Transactions on Industrial Informatics}, vol.~6, no.~4, pp.
  744--757, Nov 2010.

\bibitem{Miller:2012:SSC:2380790.2380805}
\BIBentryALTinterwordspacing
B.~Miller and D.~Rowe, ``A survey scada of and critical infrastructure
  incidents,'' in \emph{Proceedings of the 1st Annual Conference on Research in
  Information Technology}, ser. RIIT '12.\hskip 1em plus 0.5em minus
  0.4em\relax New York, NY, USA: ACM, 2012, pp. 51--56. [Online]. Available:
  \url{http://doi.acm.org/10.1145/2380790.2380805}
\BIBentrySTDinterwordspacing

\bibitem{ponemon-report}
\BIBentryALTinterwordspacing
P.~I. LLC, ``Critical infrastructure: Security preparedness and maturity,''
  2014. [Online]. Available:
  \url{https://www.hunton.com/files/upload/Unisys_Report_Critical_Infrastructure_Cybersecurity.pdf}
\BIBentrySTDinterwordspacing

\bibitem{stuxnet}
R.~Langner, ``Stuxnet: Dissecting a cyberwarfare weapon,'' \emph{IEEE Security
  Privacy}, vol.~9, no.~3, pp. 49--51, May 2011.

\bibitem{urciuoli2013supply}
L.~Urciuoli, T.~M{\"a}nnist{\"o}, J.~Hintsa, and T.~Khan, ``Supply chain cyber
  security-potential threats,'' \emph{Information \& Security}, vol.~29, no.~1,
  p.~51, 2013.

\bibitem{mitchell2014survey}
R.~Mitchell and I.-R. Chen, ``A survey of intrusion detection techniques for
  cyber-physical systems,'' \emph{ACM Computing Surveys (CSUR)}, vol.~46,
  no.~4, p.~55, 2014.

\bibitem{cheung2007using}
S.~Cheung, B.~Dutertre, M.~Fong, U.~Lindqvist, K.~Skinner, and A.~Valdes,
  ``Using model-based intrusion detection for scada networks,'' in
  \emph{Proceedings of the SCADA security scientific symposium}, vol.~46, 2007,
  pp. 1--12.

\bibitem{6016202}
Y.~Mo, T.~H.~J. Kim, K.~Brancik, D.~Dickinson, H.~Lee, A.~Perrig, and
  B.~Sinopoli, ``Cyber-physical security of a smart grid infrastructure,''
  \emph{Proceedings of the IEEE}, vol. 100, no.~1, pp. 195--209, Jan 2012.

\bibitem{6141833}
Y.~Yan, Y.~Qian, H.~Sharif, and D.~Tipper, ``A survey on cyber security for
  smart grid communications,'' \emph{IEEE Communications Surveys Tutorials},
  vol.~14, no.~4, pp. 998--1010, Fourth 2012.

\bibitem{6257525}
X.~Li, X.~Liang, R.~Lu, X.~Shen, X.~Lin, and H.~Zhu, ``Securing smart grid:
  cyber attacks, countermeasures, and challenges,'' \emph{IEEE Communications
  Magazine}, vol.~50, no.~8, pp. 38--45, August 2012.

\bibitem{wang2013cyber}
W.~Wang and Z.~Lu, ``Cyber security in the smart grid: Survey and challenges,''
  \emph{Computer Networks}, vol.~57, no.~5, pp. 1344--1371, 2013.

\bibitem{simon2017critical}
\BIBentryALTinterwordspacing
T.~Simon, ``Critical infrastructure and the internet of things,'' 2017.
  [Online]. Available:
  \url{https://www.cigionline.org/sites/default/files/documents/GCIG%20no.46_0.pdf}
\BIBentrySTDinterwordspacing

\bibitem{7959707}
P.~L. Gallegos-Segovia, J.~F. Bravo-Torres, J.~J. Argudo-Parra, E.~J.
  Sacoto-Cabrera, and V.~M. Larios-Rosillo, ``Internet of things as an attack
  vector to critical infrastructures of cities,'' in \emph{2017 International
  Caribbean Conference on Devices, Circuits and Systems (ICCDCS)}, June 2017,
  pp. 117--120.

\bibitem{BEKARA2014532}
\BIBentryALTinterwordspacing
C.~Bekara, ``Security issues and challenges for the iot-based smart grid,''
  \emph{Procedia Computer Science}, vol.~34, pp. 532 -- 537, 2014, the 9th
  International Conference on Future Networks and Communications (FNC'14)/The
  11th International Conference on Mobile Systems and Pervasive Computing
  (MobiSPC'14)/Affiliated Workshops. [Online]. Available:
  \url{http://www.sciencedirect.com/science/article/pii/S1877050914009193}
\BIBentrySTDinterwordspacing

\bibitem{sajid2016cloud}
A.~Sajid, H.~Abbas, and K.~Saleem, ``Cloud-assisted iot-based scada systems
  security: A review of the state of the art and future challenges,''
  \emph{IEEE Access}, vol.~4, pp. 1375--1384, 2016.

\bibitem{kuhrer2014hell}
M.~K{\"u}hrer, T.~Hupperich, C.~Rossow, and T.~Holz, ``Hell of a handshake:
  Abusing tcp for reflective amplification ddos attacks.'' in \emph{WOOT},
  2014.

\bibitem{sekar2006lads}
V.~Sekar, N.~G. Duffield, O.~Spatscheck, J.~E. van~der Merwe, and H.~Zhang,
  ``Lads: Large-scale automated ddos detection system.'' in \emph{USENIX Annual
  Technical Conference, General Track}, 2006, pp. 171--184.

\bibitem{fayaz2015bohatei}
S.~K. Fayaz, Y.~Tobioka, V.~Sekar, and M.~Bailey, ``Bohatei: Flexible and
  elastic ddos defense.'' in \emph{USENIX Security Symposium}, 2015, pp.
  817--832.

\bibitem{zargar2013survey}
S.~T. Zargar, J.~Joshi, and D.~Tipper, ``A survey of defense mechanisms against
  distributed denial of service (ddos) flooding attacks,'' \emph{IEEE
  communications surveys \& tutorials}, vol.~15, no.~4, pp. 2046--2069, 2013.

\bibitem{janus2011heads}
M.~Janus, ``Heads of the hydra. malware for network devices,'' 2011.

\bibitem{de2017analysis}
M.~De~Donno, N.~Dragoni, A.~Giaretta, and A.~Spognardi, ``Analysis of
  ddos-capable iot malwares,'' in \emph{Proceedings of 1st International
  Conference on Security, Privacy, and Trust (INSERT)}, 2017.

\bibitem{203628}
\BIBentryALTinterwordspacing
M.~Antonakakis, T.~April, M.~Bailey, M.~Bernhard, E.~Bursztein, J.~Cochran,
  Z.~Durumeric, J.~A. Halderman, L.~Invernizzi, M.~Kallitsis, D.~Kumar,
  C.~Lever, Z.~Ma, J.~Mason, D.~Menscher, C.~Seaman, N.~Sullivan, K.~Thomas,
  and Y.~Zhou, ``Understanding the mirai botnet,'' in \emph{26th {USENIX}
  Security Symposium ({USENIX} Security 17)}.\hskip 1em plus 0.5em minus
  0.4em\relax Vancouver, BC: {USENIX} Association, 2017, pp. 1093--1110.
  [Online]. Available:
  \url{https://www.usenix.org/conference/usenixsecurity17/technical-sessions/presentation/antonakakis}
\BIBentrySTDinterwordspacing

\bibitem{7971869}
C.~Kolias, G.~Kambourakis, A.~Stavrou, and J.~Voas, ``Ddos in the iot: Mirai
  and other botnets,'' \emph{Computer}, vol.~50, no.~7, pp. 80--84, 2017.

\bibitem{guri20179}
M.~Guri, Y.~Mirsky, and Y.~Elovici, ``9-1-1 ddos: Attacks, analysis and
  mitigation,'' in \emph{Security and Privacy (EuroS\&P), 2017 IEEE European
  Symposium on}.\hskip 1em plus 0.5em minus 0.4em\relax IEEE, 2017, pp.
  218--232.

\bibitem{burger-king-against-google-home}
\BIBentryALTinterwordspacing
J.~Kastrenakes, ``Burger king's new ad forces google home to advertise the
  whopper,'' 2017. [Online]. Available:
  \url{https://www.theverge.com/2017/4/12/15259400/burger-king-google-home-ad-wikipedia}
\BIBentrySTDinterwordspacing

\bibitem{alexa-order-dollhouse}
\BIBentryALTinterwordspacing
A.~Liptak, ``Amazon's alexa started ordering people dollhouses after hearing
  its name on tv,'' 2017. [Online]. Available:
  \url{https://www.theverge.com/2017/1/7/14200210/amazon-alexa-tech-news-anchor-order-dollhouse}
\BIBentrySTDinterwordspacing

\bibitem{smart-irrigation-systems-market}
\BIBentryALTinterwordspacing
acutemarketreports, ``smart irrigation systems market.'' [Online]. Available:
  \url{http://www.acutemarketreports.com/report/smart-irrigation-systems-market}
\BIBentrySTDinterwordspacing

\bibitem{mutchek2014moving}
M.~Mutchek and E.~Williams, ``Moving towards sustainable and resilient smart
  water grids,'' \emph{Challenges}, vol.~5, no.~1, pp. 123--137, 2014.

\bibitem{YE2016361}
\BIBentryALTinterwordspacing
Y.~Ye, L.~Liang, H.~Zhao, and Y.~Jiang, ``The system architecture of smart
  water grid for water security,'' \emph{Procedia Engineering}, vol. 154, pp.
  361 -- 368, 2016, 12th International Conference on Hydroinformatics (HIC
  2016) - Smart Water for the Future. [Online]. Available:
  \url{http://www.sciencedirect.com/science/article/pii/S1877705816318811}
\BIBentrySTDinterwordspacing

\bibitem{doi:10.1080/19443994.2014.917887}
\BIBentryALTinterwordspacing
S.~W. Lee, S.~Sarp, D.~J. Jeon, and J.~H. Kim, ``Smart water grid: the future
  water management platform,'' \emph{Desalination and Water Treatment},
  vol.~55, no.~2, pp. 339--346, 2015. [Online]. Available:
  \url{https://doi.org/10.1080/19443994.2014.917887}
\BIBentrySTDinterwordspacing

\bibitem{Smart-Water-Grid-Market}
\BIBentryALTinterwordspacing
``Smart water grid market.'' [Online]. Available:
  \url{https://www.futuremarketinsights.com/reports/smart-water-grid-market}
\BIBentrySTDinterwordspacing

\bibitem{NOAA}
\BIBentryALTinterwordspacing
``National oceanic and atmospheric administration.'' [Online]. Available:
  \url{http://www.noaa.gov/}
\BIBentrySTDinterwordspacing

\bibitem{Metno}
\BIBentryALTinterwordspacing
``Meteorologisk institutt.'' [Online]. Available: \url{https://www.met.no/}
\BIBentrySTDinterwordspacing

\bibitem{wunderground}
\BIBentryALTinterwordspacing
``Weather underground: Weather forecast and reports - long range.'' [Online].
  Available: \url{https://www.wunderground.com/}
\BIBentrySTDinterwordspacing

\bibitem{FAWN}
\BIBentryALTinterwordspacing
``Florida automated weather service.'' [Online]. Available:
  \url{https://fawn.ifas.ufl.edu/tools/}
\BIBentrySTDinterwordspacing

\bibitem{CIMIS}
\BIBentryALTinterwordspacing
``California irrigation management information system.'' [Online]. Available:
  \url{https://cimis.water.ca.gov/}
\BIBentrySTDinterwordspacing

\bibitem{DarkSky}
\BIBentryALTinterwordspacing
``Dark sky.'' [Online]. Available: \url{https://darksky.net/dev}
\BIBentrySTDinterwordspacing

\bibitem{PwsWeather}
\BIBentryALTinterwordspacing
``Pws weather.'' [Online]. Available: \url{https://www.pwsweather.com/}
\BIBentrySTDinterwordspacing

\bibitem{top-sprinklers-1}
\BIBentryALTinterwordspacing
E.~Wiki, ``The 10 best smart sprinkler systems.'' [Online]. Available:
  \url{https://wiki.ezvid.com/best-smart-sprinkler-systems}
\BIBentrySTDinterwordspacing

\bibitem{top-sprinklers-2}
\BIBentryALTinterwordspacing
postscapes, ``Top smart irrigation sprinkler controllers.'' [Online].
  Available: \url{https://www.postscapes.com/smart-irrigation-controllers/}
\BIBentrySTDinterwordspacing

\bibitem{36-Most-Water-Stressed-Countries}
W.~R. Institute, ``World 36 most water stressed countries,''
  \url{http://www.wri.org/blog/2013/12/world%E2%80%99s-36-most-water-stressed-countries}.

\bibitem{water-tariff}
\BIBentryALTinterwordspacing
G.~W. Intelligence, ``Water tariffs in major cities growing at twice the global
  inflation rate in 2017,'' 2017. [Online]. Available:
  \url{https://www.globalwaterintel.com/water-tariffs-in-major-cities-growing-at-twice-the-global-inflation-rate-in-2017}
\BIBentrySTDinterwordspacing

\bibitem{botnets-for-rent-1}
\BIBentryALTinterwordspacing
BleepingComputer, ``You can now rent a mirai botnet of 400,000 bots,'' 2016.
  [Online]. Available:
  \url{https://www.bleepingcomputer.com/news/security/you-can-now-rent-a-mirai-botnet-of-400-000-bots/}
\BIBentrySTDinterwordspacing

\bibitem{botnets-for-rent-2}
\BIBentryALTinterwordspacing
zingbox, ``Botnet-as-a-service is for sale this cyber monday!'' 2016. [Online].
  Available:
  \url{https://www.zingbox.com/blog/botnet-as-a-service-is-for-sale-this-cyber-monday/}
\BIBentrySTDinterwordspacing

\bibitem{zeidanloo2009botnet}
H.~R. Zeidanloo and A.~A. Manaf, ``Botnet command and control mechanisms,'' in
  \emph{Computer and Electrical Engineering, 2009. ICCEE'09. Second
  International Conference on}, vol.~1.\hskip 1em plus 0.5em minus 0.4em\relax
  IEEE, 2009, pp. 564--568.

\bibitem{DBLP:conf/cardis/ShwartzMBEO17}
\BIBentryALTinterwordspacing
O.~Shwartz, Y.~Mathov, M.~Bohadana, Y.~Elovici, and Y.~Oren, ``Opening
  pandora's box: Effective techniques for reverse engineering iot devices,'' in
  \emph{Smart Card Research and Advanced Applications - 16th International
  Conference, {CARDIS} 2017, Lugano, Switzerland, November 13-15, 2017, Revised
  Selected Papers}, 2017, pp. 1--21. [Online]. Available:
  \url{https://doi.org/10.1007/978-3-319-75208-2_1}
\BIBentrySTDinterwordspacing

\bibitem{Meidan:2017:PML:3019612.3019878}
\BIBentryALTinterwordspacing
Y.~Meidan, M.~Bohadana, A.~Shabtai, J.~D. Guarnizo, M.~Ochoa, N.~O.
  Tippenhauer, and Y.~Elovici, ``Profiliot: A machine learning approach for iot
  device identification based on network traffic analysis,'' in
  \emph{Proceedings of the Symposium on Applied Computing}, ser. SAC '17.\hskip
  1em plus 0.5em minus 0.4em\relax New York, NY, USA: ACM, 2017, pp. 506--509.
  [Online]. Available: \url{http://doi.acm.org/10.1145/3019612.3019878}
\BIBentrySTDinterwordspacing

\bibitem{Falcon}
\BIBentryALTinterwordspacing
``Falcon 6504 rotors.'' [Online]. Available:
  \url{http://www.rainbird.com/sites/default/files/media/documents/2018-02/ts_Falcon_6504.pdf}
\BIBentrySTDinterwordspacing

\bibitem{water-tower}
\BIBentryALTinterwordspacing
M.~BRAIN, ``How water towers work.'' [Online]. Available:
  \url{https://people.howstuffworks.com/water.htm}
\BIBentrySTDinterwordspacing

\bibitem{betarim}
\BIBentryALTinterwordspacing
``Kkl-jnf reservoirs provide priceless benefits.'' [Online]. Available:
  \url{http://www.kkl-jnf.org/water-for-israel/water-reservoirs/}
\BIBentrySTDinterwordspacing

\bibitem{DBLP:journals/corr/abs-1708-09537}
\BIBentryALTinterwordspacing
G.~Zhang, C.~Yan, X.~Ji, T.~Zhang, T.~Zhang, and W.~Xu, ``Dolphinatack:
  Inaudible voice commands,'' \emph{CoRR}, vol. abs/1708.09537, 2017. [Online].
  Available: \url{http://arxiv.org/abs/1708.09537}
\BIBentrySTDinterwordspacing

\end{thebibliography}



\footnotesize 
\Urlmuskip=0mu plus 1mu\relax

\end{document}